\documentclass[aip, pof, amsmath,amssymb, preprint, floatfix]{revtex4-1}

\usepackage{multirow}
\usepackage{array}

\usepackage{graphicx}
\usepackage{dcolumn}
\usepackage{bm}

\usepackage[utf8]{inputenc}
\usepackage[T1]{fontenc}
\usepackage{mathptmx}

\graphicspath{{Fig/}}

\begin{document}
\title{Start-up and cessation of steady shear and extensional flows: Exact analytical solutions for the affine linear Phan-Thien--Tanner fluid model}

\author{D. Shogin}
 \altaffiliation[Also at ]{The National IOR Centre of Norway, University of Stavanger, 4036 Ullandhaug, Norway} 
 \email{dmitry.shogin@uis.no}
\affiliation{Department of Energy Resources, University of Stavanger, 4036 Ullandhaug, Norway}

\date{\today}

\begin{abstract}
Exact analytical solutions for start-up and cessation flows are obtained for the affine linear Phan-Thien--Tanner fluid model. They include the results for start-up and cessation of steady shear flows, of steady uniaxial and biaxial extensional flows, and of steady planar extensional flows. The solutions obtained show that at start-up of steady shear flows, the stresses go through quasi-periodic exponentially damped oscillations while approaching their steady-flow values (so that stress overshoots are present); at start-up of steady extensional flows, the stresses grow monotonically, while at cessation of steady shear and extensional flows, the stresses decay quickly and non-exponentially. The steady-flow rheology of the fluid is also reviewed, the exact analytical solutions obtained in this work for steady shear and extensional flows being simpler than the alternative formulas found in the literature. The properties of steady and transient solutions, including their asymptotic behavior at low and high Weissenberg numbers, are investigated in detail. Generalization to the multimode version of the Phan-Thien--Tanner model is also discussed. Thus, this work provides a complete analytical description of the rheology of the affine linear Phan-Thien--Tanner fluid in start-up, cessation, and steady regimes of shear and extensional flows.
\end{abstract}

\maketitle

\section{
\label{Sec:Intro}
Introduction}
Transient flow experiments are of vital imporance to rheology: they help investigate the complex nonlinear nature of non-Newtonian fluids (in particular, of polymeric liquids), which is fully manifested in such flows.\cite{Bird1987a} The basic rheological experiments involving transient flows are the so-called step-rate (start-up and cessation) tests.\cite{Mezger2014} In start-up tests, the fluid originally at rest is set to motion by a suddenly applied constant shear or elongation rate, and the stress growth is monitored until the  steady-flow regime is established. In cessation experiments, the fluid undergoes steady shear or extensional flow until the shear or elongation rate is removed instantaneously; the stress decay is observed while the fluid approaches equilibrium.
\par 
Possessing an exact analytical solution describing the rheological response of the fluid in such experiments is advantageous in many aspects. In particular, it facilitates fitting the fluid model to experimental data, allows one to test the model for relevance and to investigate its features, reduces calculation time, and can serve as a benchmark for numerical solvers. At the same time, very few analytical solutions are known for realistic non-Newtonian fluid models mainly because of the overall complexity of the underlying equations.\cite{Bird1987b}
\par 
The analytical solutions for transient flows of non-Newtonian fluids obtained during the last few years, including those for large-amplitude oscillatory flows\cite{Saengow2015, Saengow2017a, Saengow2017b, Saengow2017c, Saengow2018, Poungthong2019a, Poungthong2019b, Saengow2019udlaos} and start-up of steady shear flows\cite{Saengow2019startup}, are obtained either for the Oldroyd 8-constant (O8) non-Newtonian fluid model \cite{Oldroyd1958,Oldroyd1961} or for its special cases. Despite its usefulness and generality, the O8 model is quite simplistic: it is constructed based on mathematical considerations only, and (albeit  nonlinear in the rate-of-strain tensor components) is linear in the stress tensor components. In contrast, physics-based non-Newtonian fluid models originating from kinetic or network theories\cite{Bird1987b} are often formulated through differential constitutive equations that are nonlinear in the stress tensor components.\cite{Bird1995} Important examples of such models are the finitely elongated nonlinear elastic (FENE-P) dumbbell model and its modifications,\cite{Bird1980, Bird1983, Chilcott1988, Shogin2020} the Phan-Thien--Tanner (PTT) model and its variants,\cite{Phan-Thien1977, Phan‐Thien1978, Ferras2019} the Giesekus model,\cite{Giesekus1982,Giesekus1983} and the more recent eXtended Pom-Pom (XPP) model.\cite{Verbeeten2001}
\par 
As might be expected, the literature is quite scarce when it comes to analytical and even semi-analytical results for physical non-Newtonian fluid models. Furthermore, such results are almost exclusively obtained for steady shear flows. \cite{Azaiez1996,Oliveira1999, Pinho2000, Alves2001, Oliveira2002, Shogin2017, Ribau2019}
Even for steady extensional flows, the analytical descriptions found in the literature are often incomplete (only asymptotic expressions are provided)\cite{Bird1980,Petrie1990,Azaiez1996,Tanner2003} or not “fully” analytical (the solution involves one or several parameters defined implicitly).\cite{Xue1998, Ferras2019,Shogin2020} Finally, we are aware of only two theoretical works on start-up flows involving constitutive equations that are nonlinear in stresses: the exact analytical solutions for the Giesekus model proposed by its author\cite{Giesekus1982} and the qualitative investigation of the PTT model by Missaghi and Petrie,\cite{Missaghi1982} both works being more than 35 years old. 
\par 
In this work, we obtain exact analytical expressions for the material functions of the PTT model related to start-up and cessation of shear and extensional---uniaxial, biaxial, and planar---flows. Our consideration shall be restricted to the so-called “simplified,” or “affine,” version of the linear PTT (SLPTT) model.\cite{Phan-Thien1977} Developed originally to simulate the rheological behavior of concentrated polymer solutions and low-density melts,\cite{Phan-Thien1977,Azaiez1996} the SLPTT model was later used to describe flows of human blood,\cite{CampoDeano2013, Ramiar2017,Gudino2019} chyme in the small intestine,\cite{ElNaby2009,Butt2020} and (with some modifications) pig liver and bread dough.\cite{Nasseri2004} In order to obtain the main result, we review the steady-flow material functions of the model and derive exact analytical expressions also for them, our formulations being simpler than those found in the literature. Thus, this work provides a complete description of the SLPTT fluid rheology in steady, start-up, and cessation regimes of shear and extensional flows. Furthermore, the steady-flow expressions obtained here are also valid for the FENE-P dumbbell model: in such flows, SLPTT and FENE-P models are completely equivalent; despite being quite obvious, this fact is rarely mentioned in the literature.\cite{Oliveira2004,Cruz2005,Poole2019}
\par 
This paper is organized as follows: In Sec. \ref{Sec:StartUpFlows}, we define the flows of interest and the material functions relevant to these flows. Then, we formulate the equations governing the dynamics of the stress tensor components in Sec. \ref{Sec:Formulation}. In Sec. \ref{Sec:DFormulation}, these equations are reformulated in terms of dimensionless variables. The analytical expressions for the material functions for steady and step-rate flows are discussed in Secs. \ref{Sec:SteadyFlowSolutions} and \ref{Sec:StartUpSolutions}, respectively. In Sec. \ref{Sec:Multimode}, we show how our results can be generalized to the multimode SLPTT model. Conclusions are presented in Sec. \ref{Sec:Conclusions}. Some technical details are given in Appendixes \ref{App:Bijection}--\ref{App:MainResult}, and the reader shall be referred to them when appropriate. Finally, the Wolfram Mathematica codes verifying the exact and asymptotic analytical results obtained in this work are included as the supplementary material.
\section{Material functions describing start-up, cessation, and steady shear and extensional flows}
\label{Sec:StartUpFlows}
 \par 
The important rheometric flows to be considered in this work are shear flow and three special cases---uniaxial, biaxial (sometimes called equibiaxial\cite{Petrie2006}), and planar---of general extensional flows; for a detailed review of shear and extensional flows, we refer the reader to the book by Bird \textit{et al.}\cite{Bird1987a} The general forms of the velocity field ($\boldsymbol{v}$), the rate-of-strain tensor $\left[\boldsymbol{\dot{\gamma}}=(\nabla \boldsymbol{v})+(\nabla \boldsymbol{v})^\mathrm{T}\right]$, the stress tensor ($\boldsymbol{\tau}$), and its upper-convected time (Oldroyd) derivative,
\begin{equation}
\boldsymbol{\tau}_{(1)} = \dfrac{\partial}{\partial t}\boldsymbol{\tau}+\boldsymbol{v}\cdot (\nabla \boldsymbol{\tau}) - (\nabla \boldsymbol{v})^\mathrm{T} \cdot \boldsymbol{\tau} -  \boldsymbol{\tau} \cdot (\nabla \boldsymbol{v}),
\end{equation}
for shear and extensional flows in a Cartesian frame of reference $(x_1,x_2,x_3)$ are given in Table \ref{Tab:TensorFormsForDifferentFlowTypes}. Throughout this paper, second-rank tensors, vectors, and scalar quantities are written in boldface Greek, boldface Latin, and lightface font, respectively. The sign convention of Bird \textit{et al.}\cite{Bird1987a} for the stress tensor is adopted. In addition, we are using the mathematically convenient approach allowing for a unified description of uniaxial and biaxial extensional flows---namely, positive elongation rates ($\dot{\epsilon}>0$) in row II of Table \ref{Tab:TensorFormsForDifferentFlowTypes} correspond to uniaxial extension, while choosing negative elongation rates ($\dot{\epsilon}<0$) yields biaxial extension, the other aspects being exactly the same for these two flows.\cite{Bird1980,Bird1987a} In contrast, for the planar extension case, it is sufficient to consider $\dot{\epsilon}>0$ due to the flow symmetry.
\par 
\begin{table*}[ht]
\caption{
\label{Tab:TensorFormsForDifferentFlowTypes}
The forms of the velocity field ($\boldsymbol{v}$), the rate-of-strain tensor ($\boldsymbol{\dot{\gamma}}$), the stress tensor ($\boldsymbol{\tau}$), and its Oldroyd derivative $\left[\boldsymbol{\tau}_{(1)}\right]$ for different flow types: shear flow (I), uniaxial and biaxial extensional flows (II, assuming $\dot{\epsilon}>0$ and $\dot{\epsilon}<0$, respectively), and planar extensional flow (III). The shear rate ($\dot{\gamma}$), the elongation rate ($\dot{\epsilon}$), and the stress tensor components are in general functions of time.}
\begin{center}
{\begin{ruledtabular}
\begin{tabular}{c c c c c}
& $\boldsymbol{v}$ 
& $ \boldsymbol{\dot{\gamma}} $ 
& $ \boldsymbol{\tau} $ 
& $ \boldsymbol{\tau}_{(1)}$ \\[1pt]
\hline \rule{0pt}{3em}
I
& $\renewcommand*{\arraystretch}{0.7} \begin{bmatrix} 
\dot{\gamma}x_2 \\ 0 \\ 0
\end{bmatrix} $
& $\renewcommand*{\arraystretch}{0.7} \begin{bmatrix} 
0 & \dot{\gamma} & 0 \\
\dot{\gamma} & 0 & 0 \\
0 & 0 & 0
\end{bmatrix} $ 
& $ \renewcommand*{\arraystretch}{0.7} \begin{bmatrix} 
\tau_{11} & \tau_{12} & 0 \\
\tau_{12} & \tau_{22} & 0 \\
0 & 0 & \tau_{33}
\end{bmatrix} $ 
& $\dfrac{\mathrm{d} \boldsymbol{\tau}}{\mathrm{d}t}-\renewcommand*{\arraystretch}{0.7}  \begin{bmatrix} 
2\tau_{12} & \tau_{22} & 0 \\
\tau_{22} & 0 & 0 \\
0 & 0 & 0
\end{bmatrix}\dot{\gamma}$ \\ \rule{0pt}{3em}
II
& $ \renewcommand*{\arraystretch}{0.7} \begin{bmatrix} 
 -\dot{\epsilon}x_1/2 \\
-\dot{\epsilon}x_2/2 \\
\dot{\epsilon}x_3
\end{bmatrix}$
& $ \renewcommand*{\arraystretch}{0.7} \begin{bmatrix} 
-\dot{\epsilon} & 0 & 0 \\
0 & -\dot{\epsilon} & 0 \\
0 & 0 & 2 \dot{\epsilon}
\end{bmatrix}$
& $ \renewcommand*{\arraystretch}{0.7} \begin{bmatrix} 
 \tau_{11} & 0 & 0 \\
 0 & \tau_{22} & 0 \\
 0 & 0 & \tau_{33}
 \end{bmatrix} $
& $\dfrac{\mathrm{d} \boldsymbol{\tau}}{\mathrm{d}t}+\renewcommand*{\arraystretch}{0.7}  \begin{bmatrix} 
\tau_{11} & 0 & 0 \\
0 & \tau_{22} & 0 \\
0 & 0 & -2\tau_{33}
\end{bmatrix}\dot{\epsilon}$ 
\\ \rule{0pt}{3em}
III 
& $ \renewcommand*{\arraystretch}{0.7} \begin{bmatrix} 
 -\dot{\epsilon}x_1 \\
0 \\
\dot{\epsilon}x_3
\end{bmatrix}$
& \renewcommand*{\arraystretch}{0.7} $\begin{bmatrix} 
- 2 \dot{\epsilon} & 0 & 0 \\
0 & 0 & 0 \\
0 & 0 & 2 \dot{\epsilon}
\end{bmatrix}$ 
& $\renewcommand*{\arraystretch}{0.7} \begin{bmatrix} 
\tau_{11} & 0 & 0 \\
0 & \tau_{22} & 0 \\
0 & 0 & \tau_{33}
\end{bmatrix} $
& $\dfrac{\mathrm{d} \boldsymbol{\tau}}{\mathrm{d}t}+\renewcommand*{\arraystretch}{0.7}  \begin{bmatrix} 
 2\tau_{11} & 0 & 0 \\
 0 & 0 & 0 \\
 0 & 0 & -2\tau_{33}
 \end{bmatrix}\dot{\epsilon}$
\end{tabular}
\end{ruledtabular}
}
\end{center}
\end{table*}
In steady flows, the shear rate ($\dot{\gamma}=\dot{\gamma}_0$) and the elongation rate ($\dot{\epsilon}=\dot{\epsilon}_0$)  are constants; the stress tensor components are time-independent. In contrast, in start-up ($+$) and cessation ($-$) flows, the shear and elongation rate take the form
\begin{equation}
\label{Eq:Formulation:StepRate}
\begin{bmatrix} 
\dot{\gamma}(t) \\ \dot{\epsilon}(t) \end{bmatrix} = \begin{bmatrix} 
\dot{\gamma}_0 \\ \dot{\epsilon}_0 \end{bmatrix} \Theta(\pm t),
\end{equation}
where $\dot{\gamma}_0$ and $\dot{\epsilon}_0$ are constants, while $\Theta$ is the Heaviside step-function: $\Theta(t)=1$ at $t \geq 0$ and $\Theta(t)=0$ otherwise; the stress tensor components are  functions of time.
\par 
The material functions describing the properties of the fluid in steady, start-up, and cessation regimes of shear and extensional flows are defined in Table \ref{Tab:MaterialFunctions}. Those related to shear flows are well-known, while those introduced for extensional flows need a brief discussion. As seen from Table \ref{Tab:TensorFormsForDifferentFlowTypes}, uniaxial and biaxial extensional flows are axially symmetric; hence, $\tau_{11}=\tau_{22}$. Therefore, only one normal stress difference, namely, $\tau_{33}-\tau_{11}$, needs to be specified to provide a complete rheological description of the flow. The material function associated with this stress difference is the uniaxial or biaxial extensional viscosity,  $\bar{\eta}$. In contrast, one can see that two normal stress differences need to be specified in planar extensional flows. Following Bird \textit{et al.}\cite{Bird1987a} and adopting their notations, we choose to specify $\tau_{33}-\tau_{11}$ and $\tau_{22}-\tau_{11}$ and denote the associated material functions $\bar{\eta}_1$ and $\bar{\eta}_2$, respectively. These shall be referred to as the first and the second planar extensional viscosities. One should note that alternative choices are also possible; in particular, the “cross-viscosity” encountered in the literature\cite{Petrie1990,Petrie2006} is easily identified as $\bar{\eta}_1-\bar{\eta}_2$.
\begin{table*}[ht]
\caption{
\label{Tab:MaterialFunctions}
Definitions of the material functions related to steady, start-up, and cessation regimes of different flow types: shear flow (I), uniaxial and biaxial extensional flows (II), and planar extensional flow (III).}
\begin{center}
{\begin{ruledtabular}
\begin{tabular}{c c c}
 & Steady flow regime & Start-up ($+$) and cessation ($-$) regimes \\[1pt]
\hline \rule{0pt}{4.3em}
 I &\renewcommand*{\arraystretch}{1.3} $\begin{array}{rl}
 \eta(\dot{\gamma}_0)&=-\dfrac{\tau_{12}(\dot{\gamma}_0)}{\dot{\gamma}_0} \\
 \Psi_1(\dot{\gamma}_0)&=-\dfrac{\tau_{11}(\dot{\gamma}_0)-\tau_{22}(\dot{\gamma}_0)}{\dot{\gamma}_0^2} \\
 \Psi_2(\dot{\gamma}_0)&=-\dfrac{\tau_{22}(\dot{\gamma}_0)-\tau_{33}(\dot{\gamma}_0)}{\dot{\gamma}_0^2}\end{array}$ 
 & \renewcommand*{\arraystretch}{1.3}
 $\begin{array}{rl}
 \eta^{\pm}(t,\dot{\gamma}_0)&=-\dfrac{\tau_{12}(t,\dot{\gamma}_0)}{\dot{\gamma}_0} \\
 \Psi_1^\pm(t,\dot{\gamma}_0)&=-\dfrac{\tau_{11}(t,\dot{\gamma}_0)-\tau_{22}(t,\dot{\gamma}_0)}{\dot{\gamma}_0^2} \\
 \Psi_2^\pm(t,\dot{\gamma}_0)&=-\dfrac{\tau_{22}(t,\dot{\gamma}_0)-\tau_{33}(t,\dot{\gamma}_0)}{\dot{\gamma}_0^2}\end{array}$  \\ \rule{0pt}{3em}
 II &
$ \bar{\eta}(\dot{\epsilon}_0)=-\dfrac{\tau_{33}(\dot{\epsilon}_0)-\tau_{11}(\dot{\epsilon}_0)}{\dot{\epsilon}_0}$
  & 
$ \bar{\eta}^\pm(t,\dot{\epsilon}_0)=-\dfrac{\tau_{33}(t,\dot{\epsilon}_0)-\tau_{11}(t,\dot{\epsilon}_0)}{\dot{\epsilon}_0}$ \\ \rule{0pt}{4em}
 III & \renewcommand*{\arraystretch}{1.3}
$\begin{array}{rl}
\bar{\eta}_1(\dot{\epsilon}_0)&=-\dfrac{\tau_{33}(\dot{\epsilon}_0)-\tau_{11}(\dot{\epsilon}_0)}{\dot{\epsilon}_0}\\
\bar{\eta}_2(\dot{\epsilon}_0)&=-\dfrac{\tau_{22}(\dot{\epsilon}_0)-\tau_{11}(\dot{\epsilon}_0)}{\dot{\epsilon}_0}
\end{array}$ 
& \renewcommand*{\arraystretch}{1.3}
$\begin{array}{rl}
\bar{\eta}_1^\pm(t,\dot{\epsilon}_0)&=-\dfrac{\tau_{33}(t,\dot{\epsilon}_0)-\tau_{11}(t,\dot{\epsilon}_0)}{\dot{\epsilon}_0}\\
\bar{\eta}_2^\pm(t,\dot{\epsilon}_0)&=-\dfrac{\tau_{22}(t,\dot{\epsilon}_0)-\tau_{11}(t,\dot{\epsilon}_0)}{\dot{\epsilon}_0}
\end{array}$ 
\end{tabular}
\end{ruledtabular}
}
\end{center}
\end{table*}
\par 
When it comes to the step-rate-related material functions, a convention on their names needs to be adopted. In this work, the material functions describing start-up flows shall be collectively referred to as “stress growth functions,” while those related to cessation flows are referred to as “stress relaxation functions.” Whenever a particular transient material function is to be mentioned, its symbol shall be used to avoid confusion.
\par 
Finally, one should note that
\begin{equation}
\lim_{t\to\infty}
\begin{bmatrix}
\eta^+(t,\dot{\gamma}_0) \\
\Psi_{1,2}^+(t,\dot{\gamma}_0) \\
\bar{\eta}^+(t,\dot{\epsilon}_0) \\
\bar{\eta}_{1,2}^+(t,\dot{\epsilon}_0)
\end{bmatrix} = 
\begin{bmatrix}
\eta^-(0,\dot{\gamma}_0) \\
\Psi_{1,2}^-(0,\dot{\gamma}_0) \\
\bar{\eta}^-(0,\dot{\epsilon}_0) \\
\bar{\eta}_{1,2}^-(0,\dot{\epsilon}_0)
\end{bmatrix}=
\begin{bmatrix}
\eta(\dot{\gamma}_0) \\
\Psi_{1,2}(\dot{\gamma}_0) \\
\bar{\eta}(\dot{\epsilon}_0) \\
\bar{\eta}_{1,2}(\dot{\epsilon}_0)
\end{bmatrix},
\end{equation}
as follows from the definitions of the material functions and from the nature of start-up and cessation tests.
We shall find it convenient in certain situations to normalize the stress growth and relaxation functions to their steady-flow analogs at the same shear or elongation rate. Any normalized stress growth function defined this way tends asymptotically to 1 at $t\to \infty$, and any normalized stress relaxation function equals 1 at $t=0$. In this work, the word “normalized” shall be used exclusively in this context. 
\section{
\label{Sec:Formulation}
Initial formulation of the start-up and cessation problems}
The constitutive equation for the single-mode SLPTT model can be written as
\begin{equation}
\label{Eq:Formulation:LPTTConstitutiveEquation}
\left(1-\dfrac{\epsilon \lambda}{\eta_0} \mathrm{tr} \boldsymbol{\tau}\right) \boldsymbol{\tau} + \lambda \boldsymbol{\tau}_{(1)}  =-\eta_0 \boldsymbol{\dot{\gamma}},
\end{equation}
where the three positive model parameters, $\eta_0$, $\lambda$, and $\epsilon$, are the zero-shear-rate viscosity, the time constant, and the extensional parameter, respectively.\cite{Phan-Thien1977} The latter mainly affects the fluid properties in extensional flows and is typically reported to be of the order of $10^{-2}$ to $10^{-1}$; the solutions obtained here are valid for a wider range of $\epsilon$. We therefore assume $0<\epsilon<1/4$, which holds in most practical applications. Finally, all the graphs shown in this work are plotted at $\epsilon=0.1$; this choice is made for visual purposes.
\par 
To formulate the equations governing the dynamics of stresses in step-rate tests as described by the SLPTT model, one substitutes $\boldsymbol{\dot{\gamma}}$, $\boldsymbol{\tau}$, and $\boldsymbol{\tau}_{(1)}$ for the chosen flow type (see Table \ref{Tab:TensorFormsForDifferentFlowTypes}) along with the corresponding form of the strain rate [Eq. (\ref{Eq:Formulation:StepRate})] into Eq. (\ref{Eq:Formulation:LPTTConstitutiveEquation}). Having written the resulting tensor equation componentwise, one arrives at nonlinear dynamical systems of first-order ordinary differential equations for the non-zero stress tensor components as shown below. In the following, the stresses shall be treated as functions of time only; their dependencies on $\dot{\gamma}_0$ and on the model parameters shall be considered parametric.
\subsection{
\label{SSec:Formulation:SteSh}
Start-up of steady shear flow}
Substituting $\boldsymbol{\dot{\gamma}}$, $\boldsymbol{\tau}$, and $\boldsymbol{\tau}_{(1)}$ from row I of Table \ref{Tab:TensorFormsForDifferentFlowTypes} into Eq. (\ref{Eq:Formulation:LPTTConstitutiveEquation}) and using Eq. (\ref{Eq:Formulation:StepRate}) with the positive sign chosen  yield
\begin{equation}
\label{Eq:Formulation:Sh:MatrixEq}
\lambda \dfrac{\mathrm{d}}{\mathrm{d}t} \begin{bmatrix}
\tau_{11} \\ \tau_{12} \\ \tau_{22} \\ \tau_{33}
\end{bmatrix}+
\left ( 1- \dfrac{\epsilon \lambda}{\eta_0}\mathrm{tr} \boldsymbol{\tau}\right)
\begin{bmatrix}
\tau_{11} \\ \tau_{12} \\ \tau_{22} \\ \tau_{33}
\end{bmatrix}  = 
\begin{bmatrix}
2\lambda \dot{\gamma}_0 \tau_{12} \\ \lambda \dot{\gamma}_0 \tau_{22} -\eta_0 \dot{\gamma}_0 \\ 0 \\ 0 
\end{bmatrix},
\end{equation}
with $\tau_{11}=\tau_{12}=\tau_{22}=\tau_{33}=0$ at $t=0$. It follows from the third and the fourth components of Eq. (\ref{Eq:Formulation:Sh:MatrixEq}) that $\tau_{22}=\tau_{33}=0$ identically. This reduces the number of equations to two and implies that the material functions related to the second normal stress difference, $\Psi_2(\dot{\gamma}_0)$ and $\Psi_2^+(t,\dot{\gamma}_0)$, vanish for the  SLPTT model.
\subsection{
\label{SSec:Formulation:UBEx}
Start-up of steady uniaxial and biaxial extensional flows}
Substituting $\boldsymbol{\dot{\gamma}}$, $\boldsymbol{\tau}$, and $\boldsymbol{\tau}_{(1)}$ from row II of Table \ref{Tab:TensorFormsForDifferentFlowTypes} into Eq. (\ref{Eq:Formulation:LPTTConstitutiveEquation}) having used Eq. (\ref{Eq:Formulation:StepRate}) with the positive sign lead to
\begin{equation}
\lambda \dfrac{\mathrm{d}}{\mathrm{d}t} \begin{bmatrix}
\tau_{11} \\ \tau_{22} \\ \tau_{33} 
\end{bmatrix} +
\left ( 1- \dfrac{\epsilon \lambda}{\eta_0}\mathrm{tr} \boldsymbol{\tau}\right)
\begin{bmatrix}
\tau_{11} \\ \tau_{22} \\ \tau_{33} 
\end{bmatrix} 
 + \lambda \dot{\epsilon}_0\begin{bmatrix}
\tau_{11} \\ \tau_{22} \\ -2\tau_{33} 
\end{bmatrix} = 
\begin{bmatrix}
\eta_0 \dot{\epsilon}_0 \\ \eta_0 \dot{\epsilon}_0 \\
 -2\eta_0 \dot{\epsilon}_0  
\end{bmatrix}, \label{Eq:Formulation:UBEx:MatrixEq}
\end{equation}
the initial conditions being $\tau_{11}=\tau_{22}=\tau_{33}=0$ at $t=0$. Since $\tau_{22}=\tau_{11}$ identically because of the axial symmetry of the flow, the number of equations in the system is reduced to two.
\subsection{
\label{SSec:Formulation:PlaEx}
Start-up of planar extensional flow}
Having inserted $\boldsymbol{\dot{\gamma}}$, $\boldsymbol{\tau}$, and $\boldsymbol{\tau}_{(1)}$ from row III of Table \ref{Tab:TensorFormsForDifferentFlowTypes} into Eq. (\ref{Eq:Formulation:LPTTConstitutiveEquation}) and used Eq. (\ref{Eq:Formulation:StepRate}) with the positive sign, one obtains
\begin{equation}
\lambda \dfrac{\mathrm{d}}{\mathrm{d}t} \begin{bmatrix}
\tau_{11} \\ \tau_{22} \\ \tau_{33} 
\end{bmatrix} +
\left ( 1- \dfrac{\epsilon \lambda}{\eta_0}\mathrm{tr} \boldsymbol{\tau}\right)
\begin{bmatrix}
\tau_{11} \\ \tau_{22} \\ \tau_{33} 
\end{bmatrix} + \lambda \dot{\epsilon}_0 \begin{bmatrix}
2\tau_{11} \\ 0 \\ -2\tau_{33} 
\end{bmatrix} = 
\begin{bmatrix}
2\eta_0 \dot{\epsilon}_0 \\ 0 \\
 -2\eta_0 \dot{\epsilon}_0  
\end{bmatrix}. \label{Eq:Formulation:PlaEx:MatrixEq}
\end{equation}
The initial conditions are $\tau_{11}=\tau_{22}=\tau_{33}=0$ at $t=0$. The solution of the second component of Eq.  (\ref{Eq:Formulation:PlaEx:MatrixEq}) is trivial, $\tau_{22}=0$ at any $t$. The number of independent variables in the system is therefore two.
\subsection{Cessation of steady shear and extensional flows}
Insertion of $\boldsymbol{\dot{\gamma}}$, $\boldsymbol{\tau}$, and $\boldsymbol{\tau}_{(1)}$ from Table \ref{Tab:TensorFormsForDifferentFlowTypes} into Eq. (\ref{Eq:Formulation:LPTTConstitutiveEquation}) with the negative sign chosen in Eq. (\ref{Eq:Formulation:StepRate}) results in similar systems of ordinary differential equations that are nearly identical to each other. For the flow types described in Table \ref{Tab:TensorFormsForDifferentFlowTypes}, these systems can be written as
\begin{equation}
\label{Eq:Formulation:Cessation:MatrixEq}
\lambda \dfrac{\mathrm{d}}{\mathrm{d}t}\tau_{ij}+\left ( 1- \dfrac{\epsilon \lambda}{\eta_0}\mathrm{tr} \boldsymbol{\tau}\right)\tau_{ij}=0,
\end{equation}
where $(i,j) \in \left\{(1,1),(1,2),(2,2),(3,3)\right\}$ for shear flows and $(i,j) \in \left\{(1,1),(2,2),(3,3)\right\}$ for extensional flows. The initial conditions are imposed so that at all the stresses are set to their steady-flow values at $t=0$; this leads to $\tau_{22}=\tau_{33}=0$ identically for cessation of steady shear flows [thus, $\Psi_2^-(t,\dot{\gamma}_0)=0$] and to $\tau_{22}=0$ identically for cessation of planar extensional flows. Thus, the number of equations in the systems reduces to two for all cessation flows considered in this work.
\section{
\label{Sec:DFormulation}
Dimensionless formulation}
Prior to solving the remaining equations of systems  (\ref{Eq:Formulation:Sh:MatrixEq})--(\ref{Eq:Formulation:Cessation:MatrixEq}), we put these equations  into the dimensionless form. Regardless of the flow type and regime, we replace the time variable, $t$, by a dimensionless one,
\begin{equation}
\label{Eq:DFormulation:DimensionlessTime}
\bar{t} = t/\lambda,
\end{equation}
so that for any time-dependent physical quantity, $\Lambda$,
\begin{equation}
\label{Eq:DFormulation:TimeDifferentiationRule}
\dfrac{\mathrm{d} \Lambda}{\mathrm{d} t} =\dfrac{1}{\lambda} \dfrac{\mathrm{d} \Lambda}{\mathrm{d} \bar{t}} = \dfrac{1}{\lambda}\Lambda',
\end{equation}
where the prime denotes differentiation with respect to $\bar{t}$. Then, we introduce the Weissenberg number, $\mathrm{Wi}$, and the  dimensionless stress combinations, $\mathbb{N}^\pm_1$, $\mathbb{N}^\pm_2$, $\mathbb{S}^\pm$, and $\mathbb{T}^\pm$, in start-up ($+$) and cessation ($-$) flows. The definitions of these quantities depend on the flow type and are given in Table \ref{Tab:DimensionlessVariables}. According to these definitions, none of the quantities $\mathrm{Wi}$, $\mathbb{N}^\pm_1$, $\mathbb{N}^\pm_2$, $\mathbb{S}^\pm$, and $\mathbb{T}^\pm$ can take negative values; this feature shall prove to be very useful. For the rest of this paper, $\mathbb{N}_1^\pm$, $\mathbb{N}_2^\pm$, $\mathbb{S}^\pm$, and $\mathbb{T}^\pm$ shall be treated as functions of $\bar{t}$ with $\epsilon$ and $\mathrm{Wi}$ as parameters.
\begin{table*}[htp]
\caption{
\label{Tab:DimensionlessVariables}
Definitions of the Weissenberg number and the dimensionless stress combinations in start-up ($+$) and cessation ($-$) of steady shear flow (I), of steady uniaxial and biaxial extensional flows (II, with $\dot{\epsilon}_0>0$ and $\dot{\epsilon}_0<0$, respectively), and of steady planar extensional flow (III).}
\begin{center}
\begin{ruledtabular}
\begin{tabular}{c c c}
I & II & III \\ 
\hline \rule{0pt}{1.2em}
$\mathrm{Wi} = \lambda \dot{\gamma}_0$ 
& $\mathrm{Wi} = \lambda \left \vert \dot{\epsilon}_0 \right \vert$ 
& $\mathrm{Wi} = \lambda \dot{\epsilon}_0$ \\ \rule{0pt}{1.8em}
$\begin{bmatrix}
\mathbb{N}_1^\pm \\
\mathbb{S^\pm}
\end{bmatrix}=-\dfrac{\epsilon}{\eta_0 \dot{\gamma}_0}
\begin{bmatrix}
\tau_{11} \\
\tau_{12}
\end{bmatrix}$
& $ \renewcommand*{\arraystretch}{1.5} \begin{bmatrix}
\mathbb{N}_1^\pm \\
\mathbb{T^\pm}
\end{bmatrix}=-\dfrac{\epsilon}{\eta_0}
\begin{bmatrix}
\dfrac{\tau_{33}-\tau_{11}}{\dot{\epsilon}_0} \\
\dfrac{\tau_{11}+2\tau_{33}}{\left \vert  \dot{\epsilon}_0 \right \vert}
\end{bmatrix}$
& $\begin{bmatrix}
\mathbb{N}_1^\pm \\
\mathbb{N}_2^\pm \\
\mathbb{T^\pm}
\end{bmatrix}=-\dfrac{\epsilon}{\eta_0 \dot{\epsilon}_0}
\begin{bmatrix}
\tau_{33}-\tau_{11} \\
\tau_{33} \\
\tau_{11}+\tau_{33}
\end{bmatrix}$ 
\end{tabular}
\end{ruledtabular}
\end{center}
\end{table*}
\par 
It is also necessary to introduce the steady-flow values of $\mathbb{N}_1^\pm$, $\mathbb{N}_2^\pm$, $\mathbb{S}^\pm$, and $\mathbb{T}^\pm$, which we denote by $\mathbb{N}_1$, $\mathbb{N}_2$, $\mathbb{S}$, and $\mathbb{T}$, respectively.  Unless otherwise stated, these quantities shall be considered  functions of $\mathrm{Wi}$ with the parameter $\epsilon$.
\par 
The material functions of the SLPTT fluid are closely related to the dimensionless stress combinations; the corresponding relations  are given in Table \ref{Tab:MatFunctionsVsDimlessVars} (vanishing material functions are not shown).
\begin{table*}[htp]
\caption{
\label{Tab:MatFunctionsVsDimlessVars}
Relations between the material functions and the dimensionless combinations of stress tensor components for start-up, cessation, and steady regimes of simple shear flow (I), of uniaxial and biaxial extensional flows (II), and of planar extensional flow (III).}
\begin{center}
\begin{ruledtabular}
\begin{tabular}{c c c}
\multicolumn{1}{c}{I} & \multicolumn{1}{c}{II} & \multicolumn{1}{c}{III} \\ 
\hline \rule{0pt}{4.5em}
$\begin{bmatrix}
\eta^\pm(\bar{t},\mathrm{Wi}) \\
\Psi_1^\pm(\bar{t},\mathrm{Wi}) \\
\eta(\mathrm{Wi}) \\
\Psi_1(\mathrm{Wi})
\end{bmatrix}=\dfrac{\eta_0}{\epsilon}
\begin{bmatrix}
\mathbb{S}^\pm \\
\lambda \mathbb{N}^\pm_1/\mathrm{Wi} \\
\mathbb{S} \\
\lambda \mathbb{N}_1/\mathrm{Wi}
\end{bmatrix}$ 
& $\begin{bmatrix}
\bar{\eta}^\pm(\bar{t},\mathrm{Wi}) \\
\bar{\eta}(\mathrm{Wi})
\end{bmatrix}=\dfrac{\eta_0}{\epsilon}
\begin{bmatrix}
\mathbb{N}^\pm_1 \\
\mathbb{N}_1
\end{bmatrix}$  
& $\begin{bmatrix}
\bar{\eta}_{1}^\pm(\bar{t},\mathrm{Wi}) \\
\bar{\eta}_{2}^\pm(\bar{t},\mathrm{Wi}) \\
\bar{\eta}_{1}(\mathrm{Wi}) \\
\bar{\eta}_{2}(\mathrm{Wi})
\end{bmatrix}=\dfrac{\eta_0}{\epsilon}
\begin{bmatrix}
\mathbb{N}^\pm_1 \\
\mathbb{N}^\pm_2 \\
\mathbb{N}_{1} \\
\mathbb{N}_2
\end{bmatrix}$ 
\end{tabular}
\end{ruledtabular}
\end{center}
\end{table*}
\subsection{
\label{SSec:DFormulation:SteSh}
Start-up of steady shear flow}
Using the dimensionless variables from column I of Table \ref{Tab:DimensionlessVariables}, one rewrites Eq. (\ref{Eq:Formulation:Sh:MatrixEq}) as
\begin{align}
(\mathbb{N}^+_1)' &= -(1 + \mathrm{Wi} \mathbb{N}_1^+) \mathbb{N}_1^+ +2 \mathrm{Wi} \mathbb{S}^+, \label{Eq:DFormulation:SteSh:1N1Evolution}\\
(\mathbb{S}^+)' &= -(1 + \mathrm{Wi} \mathbb{N}_1^+) \mathbb{S}^+ + \epsilon, \label{Eq:DFormulation:SteSh:2SEvolution}
\end{align}
with $\mathbb{N}_1^+(0)=\mathbb{S}^+(0)=0$.
\subsection{
\label{SSec:DFormulation:UBEx}
Start-up of steady uni- and biaxial extensional flow}
Inserting the dimensionless variables from column II of Table \ref{Tab:DimensionlessVariables} into Eq. (\ref{Eq:Formulation:UBEx:MatrixEq}), one gets
\begin{align}
(\mathbb{T}^+)' &= - (1 + \mathrm{Wi} \mathbb{T}^+) \mathbb{T^+} + 2 \mathrm{Wi} \mathbb{N}^+_1, \label{Eq:DFormulation:UBEx:1TEvolution}\\
(\mathbb{N}^+_1)' &= - (1 + \mathrm{Wi} \mathbb{T}^+) \mathbb{N}_1^+ + \mathrm{Wi} (\mathbb{T}^+ \pm \mathbb{N}^+_1)+3 \epsilon, \label{Eq:DFormulation:UBEx:2N1Evolution}
\end{align}
with $\mathbb{T}^+(0) = \mathbb{N}^+_1(0)=0$.
\subsection{
\label{SSec:DFormulation:PlaEx}
Start-up of steady planar extensional flow}
One rewrites Eq. (\ref{Eq:Formulation:PlaEx:MatrixEq}) in the dimensionless form using column III of Table \ref{Tab:DimensionlessVariables}. The result is
\begin{align}
(\mathbb{T}^+)' &= - (1 + \mathrm{Wi} \mathbb {T}^+) \mathbb{T}^+ + 2 \mathrm{Wi} \mathbb{N}^+_1, \label{Eq:DFormulation:PlaEx:1TEvolution}\\
(\mathbb{N}_1^+)' &= - (1 + \mathrm{Wi} \mathbb {T}^+) \mathbb{N}^+_1 +2 \mathrm{Wi} \mathbb{T}^+ + 4 \epsilon, \label{Eq:DFormulation:PlaEx:2N1Evolution}
\end{align}
with $\mathbb{T}^+(0)=\mathbb{N}^+_1(0)=0$. Note that $\mathbb{N}_2^+$ is not an independent variable: it is easy to see that
\begin{equation}
\mathbb{N}^+_2 = \dfrac{1}{2}(\mathbb{T}^++\mathbb{N}^+_1)\label{Eq:DFormulation:PlaEx:3N2Evolution}
\end{equation}
at any $\bar{t}$.
\subsection{Cessation of steady shear and extensional flows}
The dimensionless form of Eq. (\ref{Eq:Formulation:Cessation:MatrixEq}) is 
\begin{align}
(\mathbb{N}^-_1)' &= -(1+\mathrm{Wi} \mathbb{N}_1^-)\mathbb{N}_1^-, \label{Eq:DFormulation:Cessation:Shear:1N1}\\
(\mathbb{S}^-)'     &= -(1+\mathrm{Wi} \mathbb{N}_1^-)\mathbb{S}^-,\label{Eq:DFormulation:Cessation:Shear:2S}
\end{align}
with $\mathbb{N}_1^-(0)=\mathbb{N}_1$ and $\mathbb{S}^-(0)=\mathbb{S}$, for cessation of steady shear flows and
\begin{align}
(\mathbb{T}^-)'     &= -(1+\mathrm{Wi} \mathbb{T}^-)\mathbb{T}^-,\label{Eq:DFormulation:Cessation:Extension:1T}\\
(\mathbb{N}_1^-)' &= -(1+\mathrm{Wi} \mathbb{T}^-)\mathbb{N}_1^-,\label{Eq:DFormulation:Cessation:Extension:2N1}
\end{align}
with $\mathbb{T}^-(0)=\mathbb{T}$ and $\mathbb{N}_1^-(0)=\mathbb{N}_1$, for cessation of extensional flows. For $\mathbb{N}_2^-$ in cessation of planar extensional flows,
\begin{equation}
 \mathbb{N}_2^- = \dfrac{1}{2}(\mathbb{T}^-+\mathbb{N}_1^-),
 \end{equation}
which is similar to the algebraic relation (\ref{Eq:DFormulation:PlaEx:3N2Evolution}).
\par 
Combining Eqs. (\ref{Eq:DFormulation:Cessation:Shear:1N1}) and (\ref{Eq:DFormulation:Cessation:Shear:2S}) yields
\begin{equation}
\left (\dfrac{\mathbb{N}_1^-}{\mathbb{S}^-} \right)'=0,
\end{equation}
Having integrated this and used the initial conditions, one finds that
\begin{equation}
\label{Eq:DFormulation:Cessation:MatFunShear}
\dfrac{\mathbb{N}_1^-}{\mathbb{N}_1}=\dfrac{\mathbb{S}^-}{\mathbb{S}}
\end{equation}
at any $\bar{t}$ for cessation of steady shear flows. Similarly, for cessation of extensional flows,
\begin{equation}
\label{Eq:DFormulation:Cessation:MatFunExtensional}
\dfrac{\mathbb{T}^-}{\mathbb{T}}=\dfrac{\mathbb{N}_1^-}{\mathbb{N}_1}\left(=\dfrac{\mathbb{N}_2^-}{\mathbb{N}_2} \right)
\end{equation}
at any $\bar{t}$. Therefore, only one differential equation needs to be solved, namely,
\begin{equation}
\label{Eq:DFormulation:Cessation:GeneralEquation}
(\mathbb{X}^-)'=-(1+\mathrm{Wi}\mathbb{X}^-)\mathbb{X}^-,
\end{equation}
with the initial condition $\mathbb{X}^-(0)=\mathbb{X}$, where $\mathbb{X} \equiv \mathbb{N}_1$ for shear flows and $ \mathbb{X} \equiv \mathbb{T}$ for extensional flows.
\section{
\label{Sec:SteadyFlowSolutions}
Steady-flow solutions}
Prior to solving the systems of differential equations derived in Sec. \ref{Sec:DFormulation}, we shall review their algebraic steady-flow variants in this section; this needs to be done for two reasons. First, we shall see that the transient material functions describing start-up and cessation flows are readily expressed in terms of their steady-flow counterparts. Second, we already mentioned that the analytical description of steady-flow material functions of the SLPTT model found in the literature is still incomplete, especially when it comes to extensional flows; this section is also meant to fill this gap.
\par 
An important remark should be made before we proceed. The  constitutive equation of the SLPTT model [Eq. (\ref{Eq:Formulation:LPTTConstitutiveEquation})] is mathematically similar to that of the FENE-P dumbbells;\cite{Bird1980} in steady shear and extensional flows, the similarity between the models reaches complete equivalence.\cite{Oliveira2004,Cruz2005,Poole2019} Therefore, the results obtained in this section also hold for the FENE-P dumbbells, provided one makes the following parameter replacements:
\begin{align}
\eta_0 & \leftrightarrow \dfrac{b}{b+3}nk_\mathrm{B}T\lambda_H, \\
\lambda & \leftrightarrow \dfrac{b}{b+3}\lambda_H, \\ 
\epsilon & \leftrightarrow \dfrac{1}{b+3}.
\end{align}
A detailed discussion of the parameters of the FENE-P dumbbell model ($nk_\mathrm{B}T$, $\lambda_H$, and $b$) is to be found elsewhere.\cite{Bird1980,Bird1987b,Shogin2020}
\subsection{
\label{SSec:SteSol:ShearFlow}
Shear flow}
\begin{figure}[ht]
\begin{center}
\includegraphics[width=3.37in]{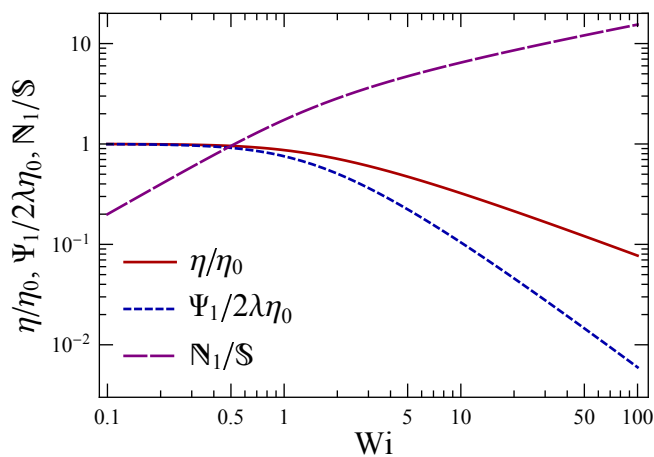}
\caption{\label{Fig:SteadyShear} Scaled non-Newtonian viscosity $\left[\eta(\mathrm{Wi})/\eta_0\right]$, scaled first normal stress coefficient $\left[\Psi_1(\mathrm{Wi})/2\lambda \eta_0\right]$, and stress ratio ($\mathbb{N}_1/\mathbb{S}$) in steady shear flows as functions of the Weissenberg number, $\mathrm{Wi}=\lambda \dot{\gamma}_0$.}
\end{center}
\end{figure}

In the steady-flow regime, Eqs. (\ref{Eq:DFormulation:SteSh:1N1Evolution}) and (\ref{Eq:DFormulation:SteSh:2SEvolution}) become
\begin{align}
(1+\mathrm{Wi} \mathbb{N}_1) \mathbb{N}_1 &= 2 \mathrm{Wi} \mathbb{S}, \label{Eq:SteSol:Sh:1FNSD}\\
(1+\mathrm{Wi} \mathbb{N}_1) \mathbb{S} &= \epsilon, \label{Eq:SteSol:Sh:2Shear}
\end{align}
respectively. Dividing Eq. (\ref{Eq:SteSol:Sh:1FNSD}) by Eq. (\ref{Eq:SteSol:Sh:2Shear}) and rearranging, one gets
\begin{equation}
\mathbb{N}_1 = \dfrac{2}{\epsilon} \mathrm{Wi} \mathbb{S}^2. \label{Eq:SteSol:Sh:Delta1SigmaRelation}
\end{equation}
Substituting this into Eq. (\ref{Eq:SteSol:Sh:2Shear}) and solving for $\mathrm{Wi}$ yield
\begin{equation}
\label{Eq:SteSol:Sh:WiSigmaRelationFinal}
\mathrm{Wi} = \sqrt{\dfrac{\epsilon(\epsilon-\mathbb{S})}{2\mathbb{S}^3}}, \quad 0<\mathbb{S} \leq \epsilon, \end{equation}
where the positive sign in front of the square root was chosen since $\mathrm{Wi} \geq 0$. The relation  (\ref{Eq:SteSol:Sh:WiSigmaRelationFinal}) is bijective (see Appendix \ref{SApp:Bijection:Wi-SigmaSteadyShearFlow}). Hence, $\mathbb{S}(\mathrm{Wi})$ is defined uniquely as the inverse of (\ref{Eq:SteSol:Sh:WiSigmaRelationFinal}). Alternatively, one can directly solve Eq. (\ref{Eq:SteSol:Sh:WiSigmaRelationFinal}) for $\mathbb{S}$, which leads to
\begin{equation}
\label{Eq:SteSol:Sh:WiSigmaRelationAlternative}
\mathbb{S} = \dfrac{1}{\mathrm{Wi}} \sqrt{\dfrac{2\epsilon}{3}}\sinh \left[\dfrac{1}{3}\mathrm{arcsinh\left(3 \sqrt{\dfrac{3\epsilon}{2}}\mathrm{Wi} \right)} \right].
\end{equation}
The solutions given by Eqs. (\ref{Eq:SteSol:Sh:WiSigmaRelationFinal}) and (\ref{Eq:SteSol:Sh:WiSigmaRelationAlternative}) are, of course, equivalent. Regardless of the one preferred, $\mathbb{N}_1(\mathrm{Wi})$ is calculated using Eq. (\ref{Eq:SteSol:Sh:Delta1SigmaRelation}). The scaled steady shear flow material functions of the SLPTT model are shown in Fig. \ref{Fig:SteadyShear} along with the stress ratio, $\mathbb{N}_1/\mathbb{S}$.

It should be noted that our analytical solution in form  (\ref{Eq:SteSol:Sh:WiSigmaRelationFinal}) can be easily used to study the asymptotic behavior of the steady shear flow material functions. In particular, it can be seen that at $\mathrm{Wi} \to 0$,
\begin{align}
\mathbb{S} &\to \epsilon, \\
\mathbb{N}_1 &\sim 2\epsilon \mathrm{Wi}, \\
\mathbb{N}_1/\mathbb{S} & \sim 2\mathrm{Wi},
\end{align} 
so that (see Table \ref{Tab:MatFunctionsVsDimlessVars})
\begin{align}
\eta & \to \eta_0, \\
\Psi_1 & \to 2 \eta_0 \lambda,
\end{align}
while at $\mathrm{Wi}\to \infty$, 
\begin{align}
\mathbb{S} &\sim \sqrt[3]{\epsilon^2/2}\mathrm{Wi}^{-2/3},\\
\mathbb{N}_1 & \sim \sqrt[3]{2\epsilon} \mathrm{Wi}^{-1/3}, \\
\mathbb{N}_1/\mathbb{S} & \sim \sqrt[3]{4/\epsilon} \mathrm{Wi}^{1/3},
\end{align}
so that (see Table \ref{Tab:MatFunctionsVsDimlessVars})
\begin{align}
\eta &\sim \eta_0 \sqrt[3]{1/2\epsilon}\mathrm{Wi}^{-2/3}, \\
\Psi_1 & \sim \eta_0 \lambda \sqrt[3]{2/\epsilon^2}\mathrm{Wi}^{-4/3}.
\end{align}
Furthermore, it follows from Eq. (\ref{Eq:SteSol:Sh:Delta1SigmaRelation}) and Table \ref{Tab:MatFunctionsVsDimlessVars} that $\Psi_1$ is proportional to the square of the viscosity,
\begin{equation}
\Psi_1=\dfrac{2\lambda}{\eta_0}\eta^2,
\end{equation}
at any $\mathrm{Wi}$.
\par 
Our results are equivalent to those found in the literature for the SLPTT model\cite{Azaiez1996,Xue1998} and for the FENE-P dumbbells\cite{Bird1980,Bird1987b,Azaiez1996,Shogin2017} but are much simpler [e.g., compare Eqs. (\ref{Eq:SteSol:Sh:WiSigmaRelationFinal}) and (\ref{Eq:SteSol:Sh:Delta1SigmaRelation}) of this work to Eqs. (37)--(40) and (48)--(51) of Azaiez \textit{et al.}\cite{Azaiez1996} or to Eqs. (16)--(19) of Xue \textit{et al.} (with $\xi=0$ and $\beta=1$)]. 
\subsection{
\label{SSec:SteSol:UBExFlow}
Uniaxial and biaxial extensional flows}
For steady flows, Eqs. (\ref{Eq:DFormulation:UBEx:1TEvolution}) and (\ref{Eq:DFormulation:UBEx:2N1Evolution}) reduce to
\begin{align}
(1+\mathrm{Wi} \mathbb{T})\mathbb{T} &= 2 \mathrm{Wi} \mathbb{N}_1, \label{Eq:SteSol:UBEx:1Trace}\\
(1+\mathrm{Wi} \mathbb{T})\mathbb{N}_1 &= \mathrm{Wi} (\mathbb{T} \pm \mathbb{N}_1) + 3 \epsilon, \label{Eq:SteSol:UBEx:2FNSD}
\end{align}
the upper and lower signs corresponding to uniaxial and biaxial extension, respectively. From Eq. (\ref{Eq:SteSol:UBEx:1Trace}),
\begin{equation}
\mathbb{N}_1 = \dfrac{(1+\mathrm{Wi} \mathbb{T})\mathbb{T}}{2\mathrm{Wi}}; \label{Eq:SteSol:UBEx:DeltaTRelation}
\end{equation}
having substituted this into Eq. (\ref{Eq:SteSol:UBEx:2FNSD}), one arrives after rearrangements at 
\begin{equation}
\label{Eq:SteSol:UBEx:WiTQuadraticEquation}
(2\mathbb{T}\pm \mathbb{T}^2-\mathbb{T}^3 )\mathrm{Wi}^2 +(6\epsilon\pm \mathbb{T}-2\mathbb{T}^2)\mathrm{Wi}-\mathbb{T} = 0.
\end{equation}
Then, Eq. (\ref{Eq:SteSol:UBEx:WiTQuadraticEquation}) is treated as quadratic in $\mathrm{Wi}$. Only one of its two solutions is physically meaningful, being non-negative and continuous at $\mathbb{T} \geq 0$. This solution can be written as
\begin{equation}
\mathrm{Wi} = \dfrac{\sqrt{36\epsilon^2\pm 12\epsilon \mathbb{T}+(9-24\epsilon) \mathbb{T}^2}-(6\epsilon \pm\mathbb{T}-2\mathbb{T}^2)}{2\mathbb{T}(2\mp \mathbb{T})(1 \pm \mathbb{T})} \label{Eq:SteSol:UBEx:WiTRelationFinal}
\end{equation}
or, with the removable singularity at $\mathbb{T}=0$ eliminated by multiplying the numerator and the denominator of Eq. (\ref{Eq:SteSol:UBEx:WiTRelationFinal}) by the conjugate of the former and simplifying the resulting fraction, as
\begin{equation}
\mathrm{Wi}=\dfrac{2\mathbb{T}}{\sqrt{36\epsilon^2\pm 12\epsilon \mathbb{T}+(9-24\epsilon) \mathbb{T}^2}+(6\epsilon \pm\mathbb{T}-2\mathbb{T}^2)}, \label{Eq:SteSol:UBEx:WiTRelationFinal2}
\end{equation}
the domain of the function being $0 \leq \mathbb{T} <2$ for uniaxial extension and $0 \leq \mathbb{T} < 1$  for biaxial extension. Equation (\ref{Eq:SteSol:UBEx:WiTRelationFinal2}) defines a bijection for both uniaxial and biaxial extension (see Appendix \ref{SApp:Bijection:Wi-TSteadyUBExFlow}). Hence, $\mathbb{T}(\mathrm{Wi})$ is defined uniquely as the inverse of (\ref{Eq:SteSol:UBEx:WiTRelationFinal}), and $\mathbb{N}_1(\mathrm{Wi})$ is  found from Eq. (\ref{Eq:SteSol:UBEx:DeltaTRelation}).
\begin{figure}[ht]
\begin{center}
\includegraphics[width=3.37in]{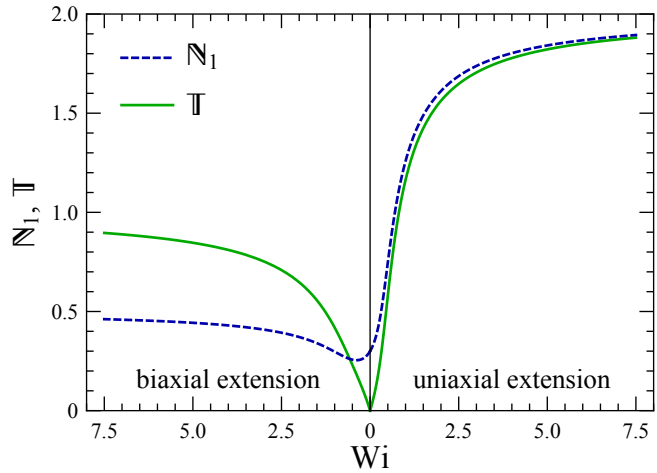}
\caption{\label{Fig:SteadyUBEx} Dimensionless stress combinations in steady uniaxial (positive elongation rates, right panel) and biaxial (negative elongation rates, left panel) extensional flows as functions of the Weissenberg number, $\mathrm{Wi}=\lambda \vert \dot{\epsilon}_0 \vert$. The shape of the $\mathbb{N}_1(\mathrm{Wi})$ curve repeats that of the extensional viscosity (see Table \ref{Tab:MatFunctionsVsDimlessVars}).}
\end{center}
\end{figure}
\par 
Functions $\mathbb{T}(\mathrm{Wi})$ and $\mathbb{N}_1(\mathrm{Wi})$ are shown in Fig. \ref{Fig:SteadyUBEx}, the right and left panels of the plot corresponding to uniaxial and biaxial extension, respectively. From Eqs. (\ref{Eq:SteSol:UBEx:WiTRelationFinal2}) and (\ref{Eq:SteSol:UBEx:DeltaTRelation}), it is seen that at $\mathrm{Wi} \to 0$,
\begin{align}
\mathbb{T} &\sim 6\epsilon \mathrm{Wi}, \\
\mathbb{N}_1 & \to 3\epsilon
\end{align}
for uniaxial and biaxial extension so that the Trouton ratio,\cite{Petrie2006}
\begin{equation}
\bar{\eta} \to 3\eta_0, \label{Eq:SteSol:UBEx:TroutonRatio}
\end{equation}
is recovered (see Table  \ref{Tab:MatFunctionsVsDimlessVars}). In both uniaxial and biaxial extensional flows, $\mathbb{T}(\mathrm{Wi})$ is increasing monotonically with $\mathrm{Wi}$ (see Appendix \ref{SApp:Bijection:Wi-TSteadyUBExFlow}) but with different asymptotic behavior at $\mathrm{Wi} \to \infty$ [see Eq. (\ref{Eq:SteSol:UBEx:WiTRelationFinal})]: in the uniaxial case,
\begin{equation}
\mathbb{T} \sim 2-\dfrac{1-\epsilon}{\mathrm{Wi}},
\end{equation}
while in the biaxial case, 
\begin{equation}
\mathbb{T} \sim 1-\dfrac{1-2\epsilon}{\mathrm{Wi}}.
\end{equation}
Furthermore, $\mathbb{N}_1(\mathrm{Wi})$, and hence the extensional viscosity, increases monotonically with $\mathrm{Wi}$ for uniaxial extension (see Appendix \ref{SApp:Viscosities:UBEx}); at $\mathrm{Wi} \to \infty$ [see Eq. (\ref{Eq:SteSol:UBEx:DeltaTRelation}) and Table \ref{Tab:MatFunctionsVsDimlessVars}],
\begin{align}
\mathbb{N}_1 & \sim 2-\dfrac{1-2\epsilon}{\mathrm{Wi}},\\
\bar{\eta} & \sim \dfrac{2\eta_0}{\epsilon}-\dfrac{(1-2\epsilon)\eta_0}{\epsilon \mathrm{Wi}}.\label{Eq:SteSol:UBEx:AsymptoticViscosityUEx}
\end{align}
In contrast, for biaxial extensional flows, $\mathbb{N}_1(\mathrm{Wi})$ and the extensional viscosity are not monotonic functions but have a minimum (see Appendix \ref{SApp:Viscosities:UBEx})---namely,
\begin{align}
\mathbb{N}_{1,\mathrm{min}} &= \dfrac{8\epsilon(1-3\epsilon)}{3-8\epsilon}, \label{Eq:SteSol:UBEx:Delta1MinimumValue}
\\
\bar{\eta}_\mathrm{min} &= \dfrac{8\eta_0(1-3\epsilon)}{3-8\epsilon}, \label{Eq:SteSol:UBEx:ExtensionalViscosityMinimumValue}
\end{align} 
at 
\begin{equation}
\label{Eq:SteSol:UBEx:ExtensionalViscosityMinimumPosition}
\mathrm{Wi}_\mathrm{min} = \dfrac{3-8\epsilon}{12(1-2\epsilon)(1-4\epsilon)};
\end{equation}
at $\mathrm{Wi} \to \infty$ [see Eq. (\ref{Eq:SteSol:UBEx:DeltaTRelation}) and Table \ref{Tab:MatFunctionsVsDimlessVars}], 
\begin{align}
\mathbb{N}_1 &\sim \dfrac{1}{2}-\dfrac{1-4\epsilon}{2\mathrm{Wi}}, \\
\bar{\eta} &\sim \dfrac{\eta_0}{2\epsilon}-\dfrac{(1-4\epsilon)\eta_0}{2\epsilon\mathrm{Wi}}. \label{Eq:SteSol:UBEx:AsymptoticViscosityBiEx}
\end{align}
\par 
We are not aware of any analogs of Eqs. (\ref{Eq:SteSol:UBEx:WiTRelationFinal2}) and (\ref{Eq:SteSol:UBEx:DeltaTRelation}) in the literature. Furthermore, to our knowledge, the character of the extensional viscosity curve (monotonic in the uniaxial case and non-monotonic in the biaxial case) was previously shown exclusively by numerical simulations and not proven analytically as in this work. Finally, we believe that we are the first to exactly describe the minimum of the biaxial extensional viscosity [see Eqs. (\ref{Eq:SteSol:UBEx:ExtensionalViscosityMinimumValue}) and (\ref{Eq:SteSol:UBEx:ExtensionalViscosityMinimumPosition})]. The asymptotic behavior of extensional viscosities at large Weissenberg numbers [Eqs. (\ref{Eq:SteSol:UBEx:AsymptoticViscosityUEx}) and (\ref{Eq:SteSol:UBEx:AsymptoticViscosityBiEx})] is in agreement with the earlier results for the SLPTT and FENE-P dumbbell fluid models \cite{Bird1980,Petrie1990,Bird1987b,Tanner2003} and thus serves as a consistency check.
\subsection{
\label{SSec:SteSol:PlaExFlow}
Planar extensional flow}
For steady planar extensional flows, Eqs. (\ref{Eq:DFormulation:PlaEx:1TEvolution}) and  (\ref{Eq:DFormulation:PlaEx:2N1Evolution}) yield
\begin{align}
(1+\mathrm{Wi} \mathbb{T})\mathbb{T} &= 2 \mathrm{Wi} \mathbb{N}_1, \label{Eq:SteSol:PlaEx:1Trace}\\
(1+\mathrm{Wi} \mathbb{T})\mathbb{N}_1 &= 2\mathrm{Wi}\mathbb{T}+4\epsilon, \label{Eq:SteSol:PlaEx:2FNSD}
\end{align}
while Eq. (\ref{Eq:DFormulation:PlaEx:3N2Evolution}) becomes
\begin{equation}
\mathbb{N}_2 = \dfrac{1}{2}(\mathbb{T}+\mathbb{N}_1). \label{Eq:SteSol:PlaEx:Delta2TDelta1Relation}
\end{equation}
Equation (\ref{Eq:SteSol:PlaEx:1Trace}) is identical in form to Eq. (\ref{Eq:SteSol:UBEx:1Trace}) for uniaxial and biaxial extensional flows; therefore,
\begin{equation}
\mathbb{N}_1 = \dfrac{(1+\mathrm{Wi} \mathbb{T})\mathbb{T}}{2\mathrm{Wi}}, \label{Eq:SteSol:PlaEx:DeltaTRelation}
\end{equation}
which is of the same form as Eq. (\ref{Eq:SteSol:UBEx:DeltaTRelation}). One proceeds by using Eq. (\ref{Eq:SteSol:PlaEx:DeltaTRelation}) to eliminate $\mathbb{N}_1$ from Eq. (\ref{Eq:SteSol:PlaEx:2FNSD}). The result is
\begin{equation}
\label{Eq:SteSol:PlaEx:WiTQuadraticEquation}
(4\mathbb{T}-\mathbb{T}^3)\mathrm{Wi}^2 + 2(4\epsilon-\mathbb{T}^2)\mathrm{Wi} - \mathbb{T} = 0.
\end{equation}
Equation (\ref{Eq:SteSol:PlaEx:WiTQuadraticEquation}) is treated as quadratic in $\mathrm{Wi}$. Of its two solutions, the one meeting the requirements of non-negativity and continuity can be written as
\begin{equation}
\mathrm{Wi} = \dfrac{2\sqrt{4\epsilon^2+(1-2\epsilon)\mathbb{T}^2}-(4\epsilon-\mathbb{T}^2) }{\mathbb{T}(4-\mathbb{T}^2)},
\label{Eq:SteSol:PlaEx:WiTRelationFinal}
\end{equation}
or, with the removable singularity at $\mathbb{T}=0$ eliminated as in Sec. \ref{SSec:SteSol:UBExFlow}, as
\begin{equation}
\mathrm{Wi} = \dfrac{\mathbb{T}}{2\sqrt{4\epsilon^2+(1-2\epsilon)\mathbb{T}^2}+(4\epsilon-\mathbb{T}^2)},
\label{Eq:SteSol:PlaEx:WiTRelationFinal2}
\end{equation}
with $0\leq \mathbb{T} <2$. Equation (\ref{Eq:SteSol:PlaEx:WiTRelationFinal2}) defines a bijection (see Appendix \ref{SApp:Bijection:Wi-TSteadyPlaExFlow}); therefore, $\mathbb{T}(\mathrm{Wi})$ is defined unambiguously as the inverse of (\ref{Eq:SteSol:PlaEx:WiTRelationFinal2}), while $\mathbb{N}_1(\mathrm{Wi})$ and $\mathbb{N}_2(\mathrm{Wi})$ are found using Eqs. (\ref{Eq:SteSol:PlaEx:DeltaTRelation}) and (\ref{Eq:SteSol:PlaEx:Delta2TDelta1Relation}), respectively.
\begin{figure}
\begin{center}
\includegraphics[width=3.37in]{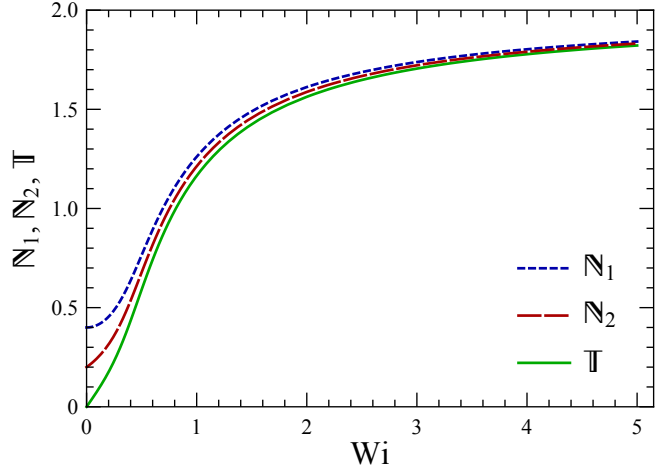}
\caption{\label{Fig:SteadyPlaEx} Dimensionless stress combinations in steady planar extensional flows as functions of the Weissenberg number, $\mathrm{Wi}=\lambda \dot{\epsilon}_0$. The shapes of the $\mathbb{N}_1(\mathrm{Wi})$ and $\mathbb{N}_2(\mathrm{Wi})$ curves repeat those of the first and the second extensional viscosity, respectively (see Table \ref{Tab:MatFunctionsVsDimlessVars}).}
\end{center}
\end{figure}
\par 
Functions $\mathbb{N}_1(\mathrm{Wi})$, $\mathbb{N}_2(\mathrm{Wi})$, and $\mathbb{T}(\mathrm{Wi})$ are shown in Fig. \ref{Fig:SteadyPlaEx}. All of them, and therefore the planar extensional viscosities, are increasing monotonically with $\mathrm{Wi}$ (see Appendix \ref{SApp:Viscosities:PlaEx} and Table \ref{Tab:MatFunctionsVsDimlessVars}). One finds that at $\mathrm{Wi} \to 0$ [see Eqs. (\ref{Eq:SteSol:PlaEx:WiTRelationFinal2}), (\ref{Eq:SteSol:PlaEx:DeltaTRelation}), and (\ref{Eq:SteSol:PlaEx:Delta2TDelta1Relation})],
\begin{align}
\mathbb{T} & \sim 8\epsilon \mathrm{Wi}, \\
\mathbb{N}_1 & \to 4\epsilon, \\
\mathbb{N}_2 & \to 2\epsilon,
\end{align}
so that (see Table \ref{Tab:MatFunctionsVsDimlessVars})
\begin{align}
\bar{\eta}_1 & \to 4\eta_0, \label{Eq:SteSol:PlaEx:AsymptoticEta1}\\
\bar{\eta}_2 & \to 2\eta_0, \label{Eq:SteSol:PlaEx:AsymptoticEta2}
\end{align}
while at $\mathrm{Wi} \to \infty$ [see Eqs. (\ref{Eq:SteSol:PlaEx:WiTRelationFinal}), (\ref{Eq:SteSol:PlaEx:DeltaTRelation}), and (\ref{Eq:SteSol:PlaEx:Delta2TDelta1Relation})],
\begin{align}
\mathbb{T}      & \sim 2-\dfrac{1-\epsilon}{\mathrm{Wi}},\\
\mathbb{N}_1& \sim 2-\dfrac{1-2\epsilon}{\mathrm{Wi}},\\
\mathbb{N}_2& \sim 2-\dfrac{2-3\epsilon}{2\mathrm{Wi}},
\end{align}
so that (see Table \ref{Tab:MatFunctionsVsDimlessVars})
\begin{align}
\bar{\eta}_1 & \sim \dfrac{2\eta_0}{\epsilon}-\dfrac{(1-2\epsilon)\eta_0}{\epsilon \mathrm{Wi}},\label{Eq:SteSol:PlaEx:AsymptoticEta1HighWi}\\
\bar{\eta}_2 & \sim \dfrac{2\eta_0}{\epsilon}-\dfrac{(2-3\epsilon)\eta_0}{2\epsilon \mathrm{Wi}}.\label{Eq:SteSol:PlaEx:AsymptoticEta2HighWi}
\end{align}
\par 
Equations (\ref{Eq:SteSol:PlaEx:WiTRelationFinal2}), (\ref{Eq:SteSol:PlaEx:DeltaTRelation}), and (\ref{Eq:SteSol:PlaEx:Delta2TDelta1Relation}) provide a complete description of rheological properties of the SLPTT fluid model in steady planar extensional flows. The only alternative to these equations we found in the literature [Eqs. (23) and (24) of Xue \textit{et al.},\cite{Xue1998} with $\xi=0$ and $\beta=1$] is both less detailed (no formula for the second planar extensional viscosity is provided) and more complicated compared to our result. The asymptotic values of the planar extensional viscosities at low and high Weissenberg numbers obtained in this work [Eqs. (\ref{Eq:SteSol:PlaEx:AsymptoticEta1}), (\ref{Eq:SteSol:PlaEx:AsymptoticEta2}),   (\ref{Eq:SteSol:PlaEx:AsymptoticEta1HighWi}), and (\ref{Eq:SteSol:PlaEx:AsymptoticEta2HighWi})] are in agreement with  general expectations\cite{Petrie2006} and with the earlier analytical results of Petrie.\cite{Petrie1990}
\section{
\label{Sec:StartUpSolutions}
Step-rate solutions}
The main result of this work (for its derivation, see Appendix \ref{App:MainResult}) is the exact analytical expressions for the stress growth and relaxation functions of the SLPTT fluid model. We have found that the normalized stress growth functions can be written in one of the two compact, closely related, and beautiful forms---those related to start-up of shear flows take the form $\mathfrak{T}$ (for “trigonometric”), while those related to start-up of extensional flows take the form $\mathfrak{H}$ (for “hyperbolic”) and the normalized stress relaxation functions related to cessation of shear and extensional flows are of the form $\mathfrak{E}$ (for “exponential”). 
\par 
Forms $\mathfrak{T}$, $\mathfrak{H}$, and $\mathfrak{E}$ are presented in Table \ref{Tab:Forms} together with the material functions taking these forms. The normalized trace of the stress tensor [$\mathbb{T}^\pm(\bar{t},\mathrm{Wi})/\mathbb{T}(\mathrm{Wi})$] is not traditionally considered a material function but can be of theoretical interest; therefore, it is also included in the results for the sake of completeness.  
\begin{table*}
\caption{\label{Tab:Forms} Trigonometric form ($\mathfrak{T}$),  hyperbolic form ($\mathfrak{H}$), exponential form ($\mathfrak{E}$), and the normalized stress growth and relaxation functions of the SLPTT fluid described by these forms.}
\begin{center}
\begin{ruledtabular}
\begin{tabular}{c c c}
Form & Material functions & Expression \\
\hline \rule{0pt}{2em}
$\mathfrak{T}$ 
& 
$\dfrac{\eta^+(\bar{t},\mathrm{Wi})}{\eta(\mathrm{Wi})}$, 
$\dfrac{\Psi_1^+(\bar{t},\mathrm{Wi})}{\Psi_1(\mathrm{Wi})}$ 
&
$1-\dfrac{K (\cos \omega \bar{t} + a\sin \omega \bar{t})}{C \exp \Omega \bar{t}+\mathrm{Wi} (A \cos \omega \bar{t} + B \sin \omega \bar{t} )}$ \\ \rule{0pt}{2em}
$\mathfrak{H}$ &
$\dfrac{\bar{\eta}^+(\bar{t},\mathrm{Wi})}{\bar{\eta}(\mathrm{Wi})}$, 
$\dfrac{\bar{\eta}_{1,2}^+(\bar{t},\mathrm{Wi})}{\bar{\eta}_{1,2}(\mathrm{Wi})}$, 
$\dfrac{\mathbb{T}^+(\bar{t},\mathrm{Wi})\footnote{at start-up of extensional flows}}{\mathbb{T}(\mathrm{Wi})}$ 
&
$1-\dfrac{K (\cosh \omega \bar{t} + a \sinh \omega \bar{t})}{C \exp \Omega \bar{t}+\mathrm{Wi} (A \cosh \omega \bar{t} + B \sinh \omega \bar{t} )}$ \\ \rule{0pt}{2em}
$\mathfrak{E}$ & 
$\dfrac{\eta^-(\bar{t},\mathrm{Wi})}{\eta(\mathrm{Wi})}$, 
$\dfrac{\Psi_1^-(\bar{t},\mathrm{Wi})}{\Psi_1(\mathrm{Wi})}$, 
$\dfrac{\bar{\eta}^-(\bar{t},\mathrm{Wi})}{\bar{\eta}(\mathrm{Wi})}$, 
$\dfrac{\bar{\eta}_{1,2}^-(\bar{t},\mathrm{Wi})}{\bar{\eta}_{1,2}(\mathrm{Wi})}$, 
$\dfrac{\mathbb{T}^-(\bar{t},\mathrm{Wi})\footnote{at cessation of extensional flows}}{\mathbb{T}(\mathrm{Wi})}$
& 
$\dfrac {1} { (1+\mathrm{Wi} \mathbb{X}) \exp\bar{t} -\mathrm{Wi} \mathbb{X}}$ 
\end{tabular}
\end{ruledtabular}
\end{center}
\end{table*}
\par 
Forms $\mathfrak{T}$ and $\mathfrak{H}$ contain expressions $\Delta>0$, $\omega = \mathrm{Wi} \sqrt{\Delta}/2$, $\Omega$, $K$, $C$, $A$, and $B$, which are functions of $\mathrm{Wi}$. The definitions of these functions depend on the flow type and are given in Table \ref{Tab:FormFunctions}. Another expression encountered in forms $\mathfrak{T}$ and $\mathfrak{H}$ is $a$, which is defined uniquely for each material function. The definitions of $a$ corresponding to different start-up material functions are found in Table \ref{Tab:a}. Finally, in the form $\mathfrak{E}$, $\mathbb{X} \equiv \mathbb{N}_1$ for cessation of shear flows and $\mathbb{X} \equiv \mathbb{T}$ for cessation of extensional flows.
\begin{table*}
\caption{
\label{Tab:FormFunctions}
Functions encountered in forms $\mathfrak{T}$ and $\mathfrak{H}$ for different flow types: start-up of steady shear flow (I), start-up of steady uniaxial and biaxial extensional flows (II, upper and lower signs, respectively), and start-up of planar extensional flow (III).}
\begin{center}
\begin{ruledtabular}
\begin{tabular}{r c c c}
 & I & II & III \\
 \hline \rule{0pt}{1.2em}
$\Delta\,$ & 
$- \mathbb{N}_1^2+8\mathbb{S} $ & 
$9-8\mathbb{N}_1 \pm 2\mathbb{T} +\mathbb{T}^2$ & 
$16-8\mathbb{N}_1+\mathbb{T}^2 $\\ \rule{0pt}{1.2em}
$\Omega\,$ &
$1 + \dfrac{3}{2} \mathrm{Wi} \mathbb{N}_1$ & 
$1 + \dfrac{1}{2}\mathrm{Wi} (\mp 1+3\mathbb{T})$ &
$1 + \dfrac{3}{2}\mathrm{Wi} \mathbb{T}$ \\ \rule{0pt}{1.2em}
$K\,$ &
$1 + \mathrm{Wi} \mathbb{N}_1 + 6 \mathrm{Wi}^2 \mathbb{S}$ & 
$1 + \mathrm{Wi} (\mp 1 + \mathbb{T}) + 2 \mathrm{Wi}^2 (-1+ 3 \mathbb{N}_1 \mp \mathbb{T})$ & 
$1 + \mathrm{Wi} \mathbb{T} + 2 \mathrm{Wi}^2 (-2+3 \mathbb{N}_1)$  \\ \rule{0pt}{1.2em}
$C\,$ &
$1 + \mathrm{Wi} \mathbb{N}_1 + 2 \mathrm{Wi}^2 \mathbb{S} $ & 
$1 + \mathrm{Wi} ( \mp 1 + \mathbb{T}) + \mathrm{Wi}^2 (-2+ 2 \mathbb{N}_1 \mp \mathbb{T} ) $ &
$1 + \mathrm{Wi} \mathbb{T} + 2 \mathrm{Wi}^2 (-2+\mathbb{N}_1)$ \\ \rule{0pt}{1.2em}
$A\,$ &
$4 \mathrm{Wi} \mathbb{S} $ & 
$\mathrm{Wi} (4\mathbb{N}_1 \mp \mathbb{T})$ &
$4 \mathrm{Wi} \mathbb{N}_1$\\ \rule{0pt}{1.2em}
$B\,$ &
$\dfrac{4 \mathbb{S}}{\sqrt{\Delta}}$ & 
$\dfrac{4\mathbb{N}_1 \mp \mathbb{T} + \mathrm{Wi}(\mp 2\mathbb{N}_1+5\mathbb{T} )} {\sqrt{\Delta}}$ & 
$\dfrac{4 (\mathbb{N}_1 + 2 \mathrm{Wi} \mathbb{T})} {\sqrt{\Delta}}$ 
\end{tabular}
\end{ruledtabular}
\end{center}
\end{table*}
\begin{table*}
\caption{
\label{Tab:a}
Function $a$ (encountered in forms $\mathfrak{T}$ and $\mathfrak{H}$) associated with different normalized stress growth functions (NMFs) in start-up of shear flow (I), of uniaxial and biaxial extensional flows (II, upper and lower signs, respectively), and of planar extensional flow (III).}
\begin{center}
\begin{ruledtabular}
\begin{tabular}{c c c c c c} 
\multicolumn{2}{c}{I}  & \multicolumn{2}{c}{II} & \multicolumn{2}{c}{III} \\ 
NMF & $a$ & NMF & $a$ & NMF & $a$ 
\\ \hline \rule{0pt}{2em}
$\dfrac{\eta^+(\bar{t},\mathrm{Wi})}{\eta(\mathrm{Wi})}$ & $-\dfrac{\mathbb{N}_1}{\sqrt{\Delta}}$ &
$\dfrac{\bar{\eta}^+(\bar{t},\mathrm{Wi})}{\bar{\eta}(\mathrm{Wi})}$  &
$\dfrac{ \pm \mathbb{N}_1 +2\mathbb{T} - \mathbb{N}_1 \mathbb{T}}{\mathbb{N}_1 \sqrt{\Delta}}$ &
$\dfrac{\bar{\eta}_1^+(\bar{t},\mathrm{Wi})}{\bar{\eta}_1(\mathrm{Wi})}$ &
$\dfrac{(4-\mathbb{N}_1)\mathbb{T}}{\mathbb{N}_1 \sqrt{\Delta}}$
\\ \rule{0pt}{2em}
$\dfrac{\Psi_1^+(\bar{t},\mathrm{Wi})}{\Psi_1(\mathrm{Wi})}$ &
$\dfrac{\mathbb{N}_1 + 2 \mathrm{Wi} \mathbb{S}}{\mathrm{Wi} \mathbb{N}_1 \sqrt{\Delta}}$ &
$\dfrac{\mathbb{T}^+(\bar{t},\mathrm{Wi})}{\mathbb{T}(\mathrm{Wi})}$ &
$\dfrac{\mathbb{T} + \mathrm{Wi} (2\mathbb{N}_1 \mp \mathbb{T})}{\mathrm{Wi} \mathbb{T} \sqrt{\Delta}}$ &
$\dfrac{\bar{\eta}_2^+(\bar{t},\mathrm{Wi})}{\bar{\eta}_2(\mathrm{Wi})}$  & 
$\dfrac {\mathbb{T} + \mathrm{Wi} (2\mathbb{N}_1 + 4\mathbb{T} -\mathbb{N}_1 \mathbb{T})} {2 \mathrm{Wi} \mathbb{N}_2 \sqrt{\Delta}}$ 
\\ \rule{0pt}{2em}
& & & &
$\dfrac{\mathbb{T}^+(\bar{t},\mathrm{Wi})}{\mathbb{T}(\mathrm{Wi})}$ &
$\dfrac{\mathbb{T} + 2\mathrm{Wi} \mathbb{N}_1}{\mathrm{Wi} \mathbb{T} \sqrt{\Delta}}$
\end{tabular}
\end{ruledtabular}
\end{center}
\end{table*}
\subsection{
\label{SSec:StartUpSolutions:ShearFlow}
Start-up of steady shear flow}
\begin{figure}
\begin{center}
\includegraphics[width=3.37in]{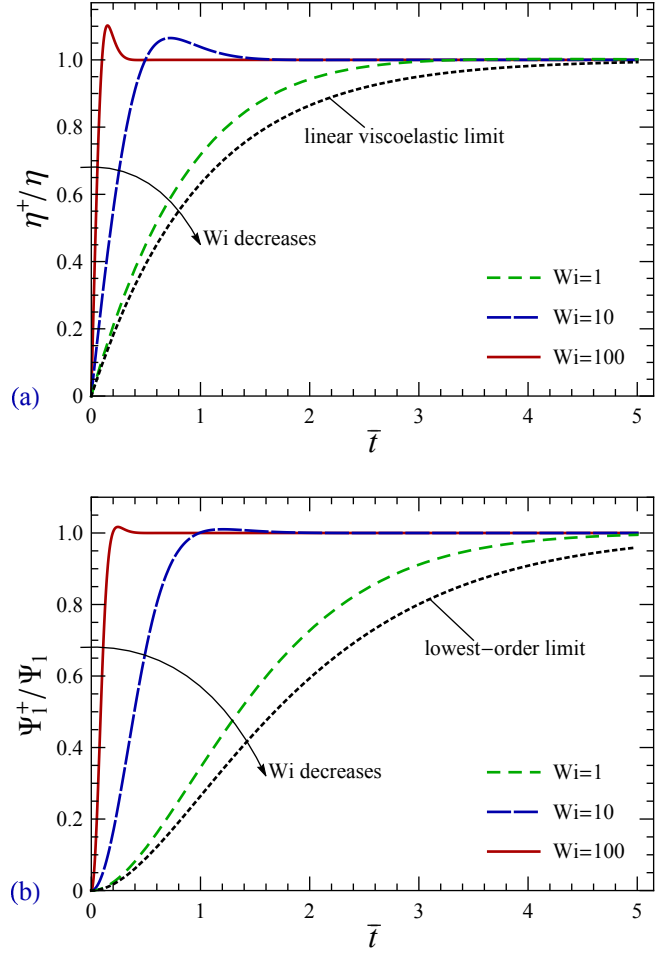}
\caption{\label{Fig:StartUpShearFlow} Normalized shear stress growth function [$\eta^+(\bar{t},\mathrm{Wi})/\eta(\mathrm{Wi})$, (a)] and first normal stress difference growth function [$\Psi_1^+(\bar{t},\mathrm{Wi})/\Psi_1(\mathrm{Wi})$, (b)] at start-up of steady shear flows as functions of the dimensionless time, $\bar{t}=t/\lambda$, at different Weissenberg numbers. The dotted curves show the limiting case $\mathrm{Wi} \to 0$.}
\end{center}
\end{figure}
The normalized stress growth functions related to start-up of steady shear flows are described by the form $\mathfrak{T}$ given in Table \ref{Tab:Forms}; the exact analytical solutions are shown in Fig. \ref{Fig:StartUpShearFlow}.  Both functions undergo quasi-periodic, exponentially damped oscillations while approaching unity.
\par 
The impact of $\mathrm{Wi}$ on these material functions is also seen in Fig. \ref{Fig:StartUpShearFlow}. At sufficiently high Weissenberg numbers, the stress overshoots become visible. As $\mathrm{Wi}$ increases, these overshoots are shifted toward earlier times and become more pronounced. At any fixed $\mathrm{Wi}$, the relative magnitude of the shear stress overshoot is always larger than that of the first normal stress difference overshoot. Finally, the steady-flow regime is approached faster at higher $\mathrm{Wi}$, with the shear stress stabilizing faster than the first normal stress difference.
\par 
In the following, we shall discuss some properties of the stress growth material functions obtained using the exact analytical solution.
\par 
First, all solutions are oscillatory, the only exception being the limiting case, $\mathrm{Wi} \to 0$. However, the oscillations are not always easy to observe because of the exponential damping [see, e.g., Figs. \ref{Fig:StartUpShearFlow}(a) and \ref{Fig:StartUpShearFlow}(b) at $\mathrm{Wi}=1$]. In fact, only the first maximum (overshoot) is well pronounced.
\par 
Second, both material functions take their steady-flow values (unity) periodically. As seen from the form $\mathfrak{T}$ (Table \ref{Tab:Forms}), it occurs when $\tan \omega \bar{t} = -1/a$. This leads to
\begin{align}
\dfrac{\eta^+}{\eta}=1 \quad \text{at} \quad & \bar{t}_k = \dfrac{1}{\omega}\arctan \dfrac{\sqrt{\Delta}}{\mathbb{N}_1}+(k-1)\dfrac{\pi}{\omega}, \label{Eq:StartUpSolutions:PeriodEtaPlus} \\
\dfrac{\Psi_1^+}{\Psi_1}=1 \quad \text{at} \quad & \bar{t}_k= -\dfrac{1}{\omega} \arctan\dfrac{\mathrm{Wi} \mathbb{N}_1 \sqrt{\Delta}}{\mathbb{N}_1+2\mathrm{Wi} \mathbb{S}}+k\dfrac{\pi}{\omega}, \label{Eq:StartUpSolutions:PeriodPsi1Plus}
\end{align}
$k$ being a natural number. Thus, the corresponding “periods,” $\Delta \bar{t}$, of $\eta^+$ and $\Psi_1^+$ are identical and equal $\pi /\omega$. Furthermore, let $\bar{t}_1$ be the time point when a normalized stress growth function reaches unity for the first time. It is seen from Eq. (\ref{Eq:StartUpSolutions:PeriodEtaPlus}) that for $\eta^+$, $0<\bar{t}_1<\pi/2\omega$. This is always smaller than $\bar{t}_1$ for $\Psi_1^+$ for which $\pi/2\omega<\bar{t}_1<\pi/\omega$ [see Eq. (\ref{Eq:StartUpSolutions:PeriodPsi1Plus})]; therefore, $\eta^+/\eta$ increases faster than $\Psi_1^+/\Psi_1$ at fixed $\mathrm{Wi}$. This can be seen from comparison of Fig. \ref{Fig:StartUpShearFlow}(a) to Fig. \ref{Fig:StartUpShearFlow}(b).
\par 
In contrast, the maxima and minima of the stress growth functions  do not occur periodically. Their positions are defined by transcendental equations (not solvable analytically) containing exponential and trigonometric functions. This can be seen by taking the time derivative of the form $\mathfrak{T}$ and setting it equal to zero.
\par 
Third, in the limit $\mathrm{Wi} \to 0$, the analytical expressions for $\eta^+$ and $\Psi_1^+$ reduce to
\begin{align}
\eta^+ & \sim \eta_0 [1-\exp(-\bar{t})], \label{Eq:StartUpSolutions:LV}\\
\Psi_1^+ & \sim 2\eta_0\lambda[1-(1+\bar{t})\exp(-\bar{t})]. \label{Eq:StartUpSolutions:LOPsi1}
\end{align}
In this limiting case [see the dotted lines in Figs. \ref{Fig:StartUpShearFlow}(a) and \ref{Fig:StartUpShearFlow}(b)], the solutions are not oscillatory. Equation (\ref{Eq:StartUpSolutions:LV}) is recognized as the so-called linear viscoelastic limit; this response, as might be expected, is identical to that of the Maxwell model.\cite{Bird1987b}
\par 
Finally, in the limit $\mathrm{Wi} \to \infty$, the form $\mathfrak{T}$ yields
\begin{align}
\dfrac{\eta^+}{\eta} & \sim 1-\dfrac{3\cos w-\sqrt{3}\sin w}{\exp(\sqrt{3}w) +2\cos w}, \label{Eq:StartUpSolutions:HighWiEtaPlus}\\
\dfrac{\Psi_1^+}{\Psi_1} & \sim  1-\dfrac{3\cos w+\sqrt{3}\sin w}{\exp(\sqrt{3}w)+2\cos w}, \label{Eq:StartUpSolutions:HighWiPsi1Plus}
\end{align}
where 
\begin{equation}
w=\sqrt[3]{\dfrac{3\sqrt{3}\epsilon}{4}}\mathrm{Wi}^{2/3}\bar{t}.
\end{equation}
Having taken the derivatives of the right-hand sides of Eqs. (\ref{Eq:StartUpSolutions:HighWiEtaPlus}) and (\ref{Eq:StartUpSolutions:HighWiPsi1Plus}) and set them equal to zero, one can find the smallest positive roots of the resulting transcendental equations numerically, e.g., using Newton's method. Then, it can be shown that
\begin{align}
\mathrm{max}(\eta^+\!/\eta) \approx 1.114 & \text{ at } \bar{t} \approx 1.468 \epsilon^{-1/3}\mathrm{Wi}^{-2/3}, \label{Eq:StartUpSolutions:OvershootEtaPlus}\\
\mathrm{max}(\Psi_1^+\!/\Psi_1) \approx 1.019 & \text{ at } \bar{t} \approx 2.394\epsilon^{-1/3}\mathrm{Wi}^{-2/3}.\label{Eq:StartUpSolutions:OvershootPsi1Plus}
\end{align}
This provides the asymptotic expressions for the positions of the first overshoots at high Weissenberg numbers. In addition, Eqs. (\ref{Eq:StartUpSolutions:OvershootEtaPlus}) and (\ref{Eq:StartUpSolutions:OvershootPsi1Plus}) set upper theoretical limits on the relative magnitudes of the stress overshoots; remarkably, these limits depend neither on $\mathrm{Wi}$ nor on the model parameters.
\subsection{
\label{SSec:StartUpSolutions:ExtensionalFlows}
Start-up of steady uniaxial, biaxial, and planar extensional flows}
\begin{figure}
\begin{center}
\includegraphics[width=3.37in]{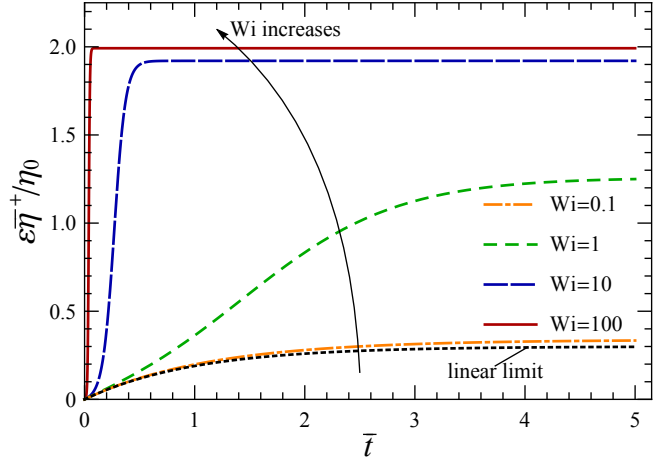}
\caption{\label{Fig:StartUpUniExFlow} Normal stress difference growth function [$\bar{\eta}^+(\bar{t},\mathrm{Wi})$] at start-up of steady uniaxial extensional flows (exact analytical solution for the SLPTT model)  plotted against the dimensionless time, $\bar{t}=t/\lambda$, at different Weissenberg numbers. The material function is scaled using a constant factor of $\epsilon/\eta_0$ for visual purposes. The dotted line shows the limit $\mathrm{Wi} \to 0$.}
\end{center}
\end{figure}

\begin{figure*}
\begin{center}
\includegraphics[width=3.37in]{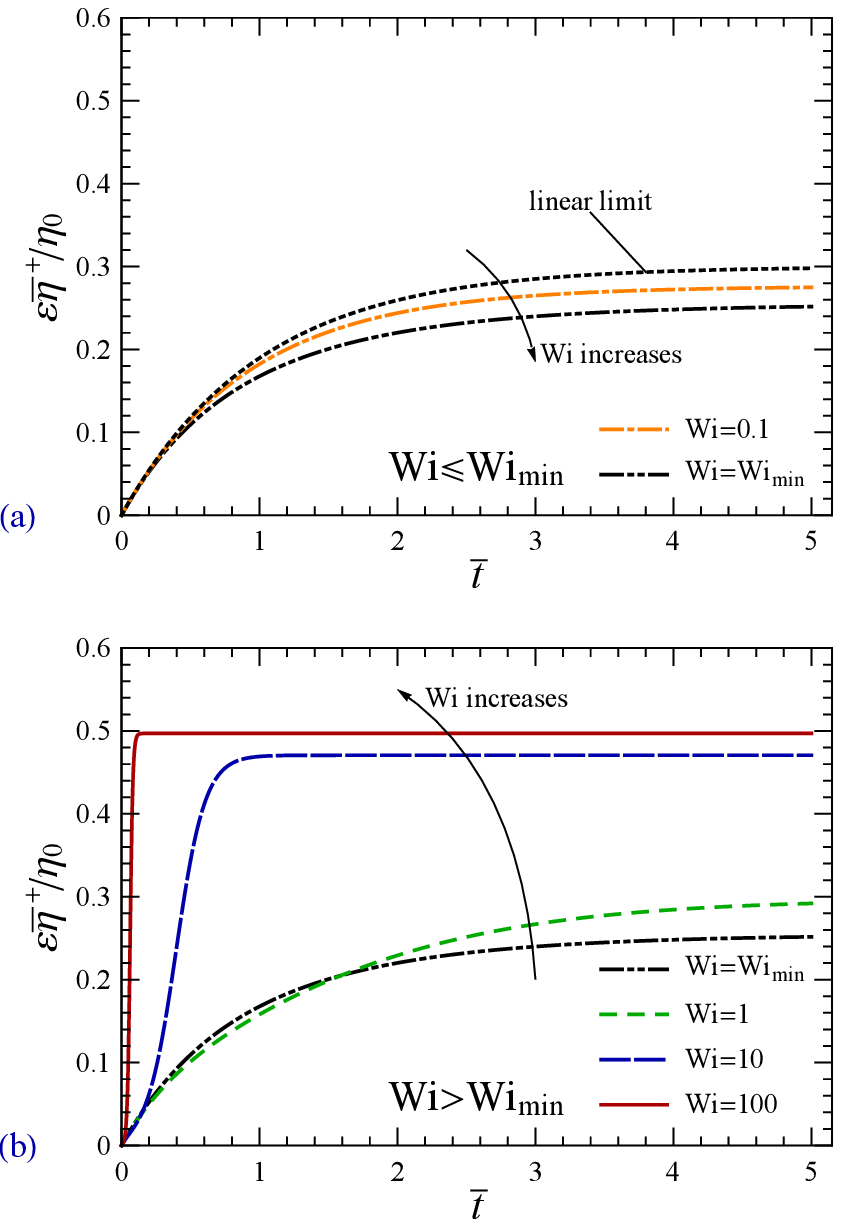}
\caption{\label{Fig:StartUpBiExFlow} Normal stress difference growth function [$\bar{\eta}^+(\bar{t},\mathrm{Wi})$] at start-up of steady biaxial extensional flows plotted against the dimensionless time, $\bar{t}=t/\lambda$, at different Weissenberg numbers: $\mathrm{Wi} \leq \mathrm{Wi}_\mathrm{min}$ (a) and $\mathrm{Wi}\geq \mathrm{Wi}_\mathrm{min}$ (b). The change in behavior observed at $\mathrm{Wi}=\mathrm{Wi}_\mathrm{min}$ [see Eq. (\ref{Eq:SteSol:UBEx:ExtensionalViscosityMinimumPosition}); at $\epsilon=0.1$ (this figure), $\mathrm{Wi}_\mathrm{min}=55/144\approx 0.38$] is due to the minimum in the extensional viscosity. The material function is scaled using a constant factor of $\epsilon/\eta_0$ for visual purposes. The dotted line shows the limit $\mathrm{Wi} \to 0$.}
\end{center}
\end{figure*}

The normalized stress growth functions related to start-up of steady extensional flows are described by the form $\mathfrak{H}$ given in Table \ref{Tab:Forms}; the exact analytical results are shown in Figs. \ref{Fig:StartUpUniExFlow} and \ref{Fig:StartUpBiExFlow} (uniaxial and biaxial extension, respectively; the results for planar extension are qualitatively similar to those for uniaxial extension). The stress growth functions approach their steady-flow values monotonically; oscillations, and therefore overshoots, are not present.
\par 
The functions $\bar{\eta}^+$, $\bar{\eta}_1^+$, and $\bar{\eta}_2^+$ are affected by the Weissenberg number in the same way: the shape of the curves changes gradually from smoother to more abrupt and step-like as $\mathrm{Wi}$ increases, the steady state being reached faster at higher $\mathrm{Wi}$.
\par 
At $\mathrm{Wi}\to 0$, the expressions for the stress growth functions reduce to
\begin{align}
\bar{\eta}^+    	& \sim 3\eta_0[1-\exp(-\bar{t})], \\
\bar{\eta}_1^+ 	& \sim 4\eta_0[1-\exp(-\bar{t})], \\
\bar{\eta}_2^+	& \sim 2\eta_0[1-\exp(-\bar{t})],
\end{align} 
the form of $\bar{\eta}^+$ for uniaxial and biaxial extension being the same in this limit.
\par 
At $\mathrm{Wi} \to \infty$,
\begin{equation}
\bar{\eta}^+ \sim \bar{\eta}_1^+ \sim \bar{\eta}_2^+ \sim \dfrac{2\eta_0}{\epsilon}\left( 1-\left[1+\dfrac{\epsilon \exp (2\mathrm{Wi} \bar{t})}{2\mathrm{Wi}} \right]^{-1}\right)
\label{Eq:StartUpExtensional:Equivalence}
\end{equation}
for start-up of steady uniaxial and planar extensional flows, while
\begin{equation}
\bar{\eta}^+ \sim \dfrac{\eta_0}{2\epsilon}\left( 1-\left[1+\dfrac{2\epsilon \exp (\mathrm{Wi}\bar{t})}{\mathrm{Wi}} \right]^{-1}\right)
\end{equation}
for start-up of steady biaxial extensional flows.
\subsection{Cessation of steady shear and extensional flows}
\begin{figure*}
\begin{center}
\includegraphics[width=3.37in]{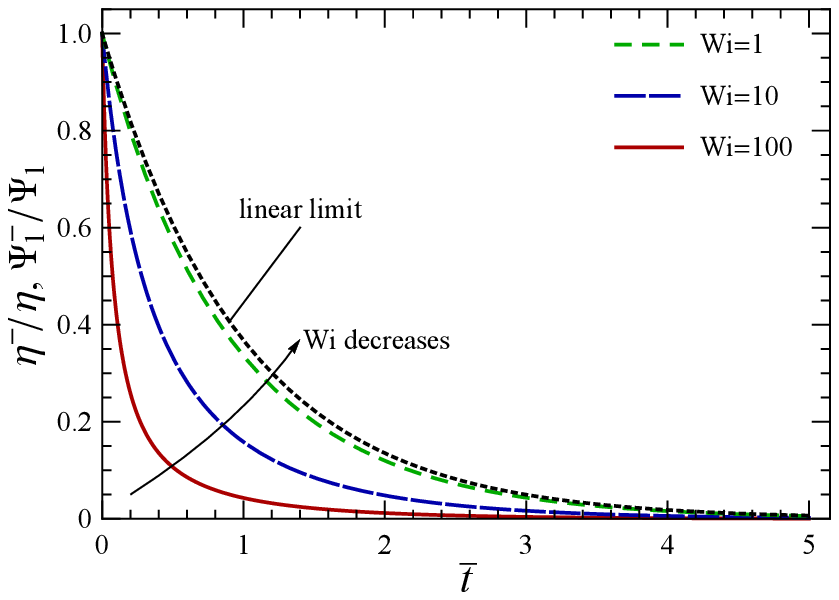}
\caption{\label{Fig:CessationShearFlow} The normalized stress relaxation functions at cessation of steady shear flow plotted against the dimensionless time, $\bar{t}=t/\lambda$, at different Weissenberg numbers. The dotted line shows the linear limit, $\mathrm{Wi} \to 0$.}
\end{center}
\end{figure*}

The normalized stress relaxation functions related to cessation of steady shear and extensional flows are described by the form $\mathfrak{E}$ in Table \ref{Tab:Forms}; the exact analytical results for shear flows (the results for extensional flows are qualitatively similar) are shown in Fig. \ref{Fig:CessationShearFlow}. The stress relaxation functions approach zero monotonically; at late times, they decay exponentially.
\par 
It is also seen that the normalized stress relaxation functions decrease faster at higher Weissenberg numbers. In the limit $\mathrm{Wi} \to 0$,
\begin{equation}
 \dfrac{\eta^-}{\eta} \sim \dfrac{\Psi_1^-}{\Psi_1} \sim \dfrac{\bar{\eta}^-}{\bar{\eta}} \sim \dfrac{\bar{\eta}_{1,2}^-}{\bar{\eta}_{1,2}}\sim \exp (-\bar{t}).
 \end{equation}
 For $\eta^-/\eta$, this result is identical to the “linear viscoelastic response” of the Maxwell model.\cite{Bird1987a}
\section{Generalization to multiple modes}
\label{Sec:Multimode}
The exact analytical solutions discussed so far were obtained for a single-mode SLPTT model. At the same time, the multimode versions of the PTT models are often used in applications to achieve a much better fit to the experimental data.\cite{Bird1987b,Hatzikiriakos1997,Langouche1999,Ramiar2017} For the SLPTT model with $M$ modes, the stress tensor, $\boldsymbol{\tau}$, is the sum of contributions from different modes,
\begin{equation}
\label{Eq:Formulation:Multimode}
\boldsymbol{\tau} = \sum_{m=1}^{M}\boldsymbol{\tau}^{(m)},
\end{equation}
where each of the contributions, $\boldsymbol{\tau}^{(m)}$, obeys the constitutive equation  (\ref{Eq:Formulation:LPTTConstitutiveEquation}) and is characterized by the corresponding $m$th mode zero-shear-rate viscosity, $\eta_0^{(m)}$, and time constant, $\lambda^{(m)}$.
\par 
The analytical results of this work for the single-mode SLPTT model are easily generalized to the case of multiple modes using Eq. (\ref{Eq:Formulation:Multimode}), the relations between the material functions and the dimensionless variables (Table \ref{Tab:MatFunctionsVsDimlessVars}), and the exact analytical solutions obtained in this work. For example, for the non-Newtonian viscosity, $\eta(\dot{\gamma}_0)$, one gets
\begin{equation}
\eta(\dot{\gamma}_0)=-\dfrac{1}{\dot{\gamma}_0}\sum_{m=1}^M\tau_{12}^{(m)}(\dot{\gamma}_0)=\dfrac{1}{\epsilon}\sum_{m=1}^M \eta_0^{(m)}\mathbb{S}\left(\lambda^{(m)}\dot{\gamma}_0\right),
\end{equation}
while the shear stress growth function, $\eta^+(t,\dot{\gamma}_0)$, is
\begin{equation}
\eta^+(t,\dot{\gamma}_0) = -\dfrac{1}{\dot{\gamma}_0}\sum_{m=1}^M\tau_{12}^{(m)}(t,\dot{\gamma}_0)=\dfrac{1}{\epsilon}\sum_{m=1}^M\mathbb{S}\left(\lambda^{(m)}\dot{\gamma}_0\right)\mathfrak{T}\left(\dfrac{t}{\lambda^{(m)}},\lambda^{(m)} \dot{\gamma}_0 \right).
\end{equation}
The exact expressions for other steady and transient material functions considered in this work can be obtained in the same way.
\section{Conclusions}
\label{Sec:Conclusions}
In this work, we have obtained and investigated the exact analytical solutions for start-up, cessation and steady-flow regimes of shear flow and of uniaxial, biaxial, and planar extensional flows having used the single-mode and multimode versions of the SLPTT fluid model.
\par 
Our most important result is the expressions for the start-up material functions (forms $\mathfrak{T}$ and $\mathfrak{H}$ in Table \ref{Tab:Forms} and the functions in these forms given in Tables \ref{Tab:FormFunctions} and \ref{Tab:a}), for which we have solved systems of coupled nonlinear differential equations. To our knowledge, we report the first (since Giesekus\cite{Giesekus1982}) exact result for the stress growth functions obtained for a physics-based non-Newtonian fluid model that is nonlinear in the stress tensor components.
\par 
The expressions for the normalized stress relaxation functions (form $\mathfrak{E}$ in Table \ref{Tab:Forms}) are much simpler to obtain; we find it surprising that they have been overlooked for many years.
\par 
Our analytical results for steady flows [Eqs. (\ref{Eq:SteSol:Sh:WiSigmaRelationAlternative}) and (\ref{Eq:SteSol:Sh:Delta1SigmaRelation}) for shear flows, Eqs. (\ref{Eq:SteSol:UBEx:WiTRelationFinal2}) and (\ref{Eq:SteSol:UBEx:DeltaTRelation}) for uniaxial and biaxial extensional flows, and Eqs. (\ref{Eq:SteSol:PlaEx:WiTRelationFinal2}), (\ref{Eq:SteSol:PlaEx:DeltaTRelation}), and (\ref{Eq:SteSol:PlaEx:Delta2TDelta1Relation}) for planar extensional flows along with Table \ref{Tab:MatFunctionsVsDimlessVars}] not only fill the existing gaps (mainly related to extensional flows) but also are significantly simpler than any of their available analogs we are aware of.
\par 
Notably, only basic knowledge of calculus and differential equations is required for understanding the mathematical methods we have used in this work. Therefore, we believe that our results are not only of scientific but also of pedagogical interest; this paper can be used when teaching theoretical rheology and non-Newtonian fluid mechanics to students.
\section*{Supplementary material}
See the supplementary material for the Wolfram Mathematica codes verifying the analytical results obtained in this work by plotting them together with the corresponding numerical solutions. These codes contain interactive plots, which can be used to see the impact of $\epsilon$ on the steady-flow material functions and that of $\epsilon$ and $\mathrm{Wi}$ on the step-rate material functions.

\begin{acknowledgments}
This research has been funded by VISTA---a basic research program in collaboration between The Norwegian Academy of Science and Letters and Equinor. D.S. is thankful to Tamara Shogina for suggestions on improvement of this paper.
\end{acknowledgments}

\begin{appendix}
\section{
\label{App:Bijection}
Bijective properties of certain functions}
\subsection{ 
\label{SApp:Bijection:Wi-SigmaSteadyShearFlow}
$\mathrm{Wi}(\mathbb{S})$ in steady shear flow}
Differentiating Eq. (\ref{Eq:SteSol:Sh:WiSigmaRelationFinal}) with respect to $\mathbb{S}$ (taking into account that $\mathbb{S}>0$) yields
\begin{equation}
\dfrac{\mathrm{d}\mathrm{Wi}}{\mathrm{d}\mathbb{S}} = \dfrac{\epsilon(2\mathbb{S}-3\epsilon)}{2\sqrt{2\epsilon(\epsilon-\mathbb{S})\mathbb{S}^5}},
\end{equation}
which is obviously negative at $0<\mathbb{S} < \epsilon$; therefore,  the function $\mathrm{Wi}(\mathbb{S})$ defined by Eq. (\ref{Eq:SteSol:Sh:WiSigmaRelationFinal}) is monotonically decreasing on its domain and thus bijective.
\subsection{
\label{SApp:Bijection:Wi-TSteadyUBExFlow}
$\mathrm{Wi}(\mathbb{T})$ in steady uniaxial and biaxial extensional flows}
Taking the derivative of Eq. (\ref{Eq:SteSol:UBEx:WiTRelationFinal2}) with respect to $\mathbb{T}$ yields
\begin{equation}
\dfrac{\mathrm{d}\mathrm{Wi}}{\mathrm{d}\mathbb{T}}=\dfrac{
4\left[3\epsilon(6\epsilon\pm \mathbb{T})+(3\epsilon+\mathbb{T}^2)Y \right]
}{
Y\left[Y+(6\epsilon \pm \mathbb{T}-2\mathbb{T}^2)\right]^2
},
\label{AEq:UBEx:DWiDTau}
\end{equation}
where
\begin{equation}
Y=\sqrt{36\epsilon^2 \pm 12\epsilon \mathbb{T}+(9-24\epsilon) \mathbb{T}^2}.
\end{equation}
\par 
For uniaxial extension, the upper signs in Eq. (\ref{AEq:UBEx:DWiDTau}) are chosen. Then, both the numerator and the denominator are clearly positive; hence, $\mathrm{Wi}(\mathbb{T})$ is a monotonically increasing function and defines a bijection.
\par 
For biaxial extension [lower signs in Eq. (\ref{AEq:UBEx:DWiDTau})], the “quick look” method is not applicable. Instead, we shall obtain the conditions at which $\mathrm{d} \mathrm{Wi} / \mathrm{d}\mathbb{T}=0$ and show that these conditions cannot be fulfilled at allowed values of $\mathbb{T}$ and $\epsilon$. This will prove that $\mathrm{Wi}(\mathbb{T})$ is a monotonic function on its domain and thus defines a bijection.
\par 
Setting the numerator of Eq. (\ref{AEq:UBEx:DWiDTau}) with the lower signs equal to zero and solving the resulting equation for $\epsilon$ lead to
\begin{align}
\epsilon^{(\pm)} &=\dfrac{\mathbb{T}^2\left[9-2\mathbb{T}(1+2\mathbb{T})\pm(1+2\mathbb{T})\sqrt{9+4\mathbb{T}^2}\right]}{12 (-2+2\mathbb{T} + 3\mathbb{T}^2)}, \nonumber \\
& \quad\;\, 0\leq \mathbb{T} <1. \label{AEq:UBEx:EpsilonPlusMinus}
\end{align}
In the following, we shall demonstrate that $\epsilon^{(-)}$ is non-positive [case (a)], while $\epsilon^{(+)}$ is either non-positive [case (b)] or exceeds $1/4$ [case (c)]. Since none of these requirements can be met (recall that $0<\epsilon<1/4$), this will complete the proof.
\par 
\textbf{Case (a).} One multiplies the numerator---with the negative sign chosen---and the denominator of Eq. (\ref{AEq:UBEx:EpsilonPlusMinus}) by the positive factor $9-2\mathbb{T}(1+2\mathbb{T})+(1+2\mathbb{T})\sqrt{9+4\mathbb{T}^2}$. After trivial algebraic manipulations, this yields
\begin{equation}
\epsilon^{(-)}=-\dfrac{3\mathbb{T}^2}{9-2\mathbb{T}(1+2\mathbb{T})+(1+2\mathbb{T})\sqrt{9+4\mathbb{T}^2}},
\end{equation}
which is clearly non-positive at $0\leq \mathbb{T} <1$.
\par 
\textbf{Case (b).} One rewrites the expression in the numerator of Eq. (\ref{AEq:UBEx:EpsilonPlusMinus}), with the positive sign chosen,
\begin{equation}
9-2\mathbb{T}(1+2\mathbb{T})+(1+2\mathbb{T})\sqrt{9+4\mathbb{T}^2}=9+(1+2\mathbb{T})(\sqrt{9+4\mathbb{T}^2}-2\mathbb{T}),
\end{equation}
which shows that the numerator is non-negative at $0\leq \mathbb{T} <1$. At the same time, the denominator changes its sign from negative to positive at $\mathbb{T}=\mathbb{T}^\ast=(\sqrt{7}-1)/3$. Therefore, $\epsilon^{(+)}$ is non-positive when $0\leq \mathbb{T} < \mathbb{T}^\ast$.
\par 
\textbf{Case (c).} As shown in case (b), $\epsilon^{(+)}$ is positive at $\mathbb{T}^\ast<\mathbb{T}<1$. However,
\begin{equation}
\epsilon^{(+)}-\dfrac{1}{4}=\dfrac{6(1-\mathbb{T})+\mathbb{T}^2(1+2\mathbb{T})(\sqrt{9+4\mathbb{T}^2}-2\mathbb{T})}{12 (-2+2\mathbb{T}+3 \mathbb{T}^2)},
\label{AEq:UBEx:EpsMinusOneFourth}
\end{equation}
which is also clearly positive; therefore, $\epsilon^{(+)}>1/4$ at $\mathbb{T}^\ast<\mathbb{T}<1$. This completes the last step of the proof.
\subsection{
\label{SApp:Bijection:Wi-TSteadyPlaExFlow}
$\mathrm{Wi}(\mathbb{T})$ in steady planar extensional flow}
Having differentiated Eq. (\ref{Eq:SteSol:PlaEx:WiTRelationFinal2}) with respect to $\mathbb{T}$, one arrives after rearrangements at
\begin{equation}
\label{AEq:PlaExProofs:DWiOverDTau}
\dfrac{\mathrm{d}\mathrm{Wi}}{\mathrm{d}\mathbb{T}}=\dfrac{
8\epsilon^2+(4\epsilon+\mathbb{T}^2)Y}
{Y(4\epsilon-\mathbb{T}^2+Y)^2},
\end{equation}
where 
\begin{equation}
Y=\sqrt{4\epsilon^2+(1-2\epsilon)\mathbb{T}^2}.
\end{equation}
It is seen that $\mathrm{d} \mathrm{Wi} / \mathrm{d}\mathbb{T}>0$; thus, Eq. (\ref{Eq:SteSol:PlaEx:WiTRelationFinal2}) specifies a monotonically increasing function and, therefore, defines a bijection.

\section{
\label{App:Viscosities}
Shapes of the extensional viscosity curves in steady flows}
\subsection{ 
\label{SApp:Viscosities:UBEx}
Monotonicity of the uniaxial extensional viscosity and existence of minimum in the biaxial extensional viscosity}
One starts with using Eq. (\ref{Eq:SteSol:UBEx:WiTRelationFinal2}) to eliminate $\mathrm{Wi}$ from Eq. (\ref{Eq:SteSol:UBEx:DeltaTRelation}). The resulting equation is then differentiated with respect to $\mathbb{T}$, which leads to
\begin{equation}
\label{AEq:Viscosities:DDeltaOneOverDWi}
\dfrac{\mathrm{d}\mathbb{N}_1}{\mathrm{d}\mathbb{T}}=
\dfrac{3(3-8\epsilon)\mathbb{T}\pm(Y+6\epsilon)}{4Y},
\end{equation}
where
\begin{equation}
\label{AEq:Viscosities:X}
Y = \sqrt{3\left[(\mathbb{T}\pm2\epsilon)^2+2\mathbb{T}^2(1-4\epsilon)+8\epsilon^2\right]}>0.
\end{equation}
For uniaxial extension [upper signs in Eqs. (\ref{AEq:Viscosities:DDeltaOneOverDWi}) and (\ref{AEq:Viscosities:X})], $\mathrm{d} \mathbb{N}_1 / \mathrm{d}\mathbb{T}$ is clearly positive (recall that $0\leq \mathbb{T}<2$ and $0<\epsilon<1/4$). For biaxial extension [lower signs in Eqs. (\ref{AEq:Viscosities:DDeltaOneOverDWi}) and (\ref{AEq:Viscosities:X})], the sign of $\mathrm{d} \mathbb{N}_1 / \mathrm{d} \mathbb{T}$ depends on the numerator of the right-hand side of Eq. (\ref{AEq:Viscosities:DDeltaOneOverDWi}). The latter changes its sign from negative to positive at
\begin{equation}
\label{AEq:UBEx:MinimumTau}
\mathbb{T} = \dfrac{4\epsilon}{3-8\epsilon},
\end{equation}
which belongs to the interval $[0,1)$ provided that $\epsilon<1/4$. 
\par 
Having applied the chain rule to $\mathbb{N}_1$, one writes
\begin{equation}
\label{AEq:UBEx:ChainRule}
\dfrac{\mathrm{d}\mathbb{N}_1}{\mathrm{d}\mathrm{Wi}}=
\dfrac{\mathrm{d}\mathbb{N}_1}{\mathrm{d}\mathbb{T}}
\dfrac{\mathrm{d}\mathbb{T}}{\mathrm{d}\mathrm{Wi}}.
\end{equation}
Since $\mathrm{d} \mathbb{T} / \mathrm{d}\mathrm{Wi}>0$, the sign of $\mathrm{d} \mathbb{N}_1 / \mathrm{d} \mathrm{Wi}$ is identical to that of $\mathrm{d} \mathbb{N}_1 / \mathrm{d}\mathbb{T}$. Therefore, $\mathbb{N}_1(\mathrm{Wi})$, and hence the extensional viscosity, is increasing monotonically in uniaxial extensional flows; while for biaxial extensional flows, it goes  through a minimum point. Substituting Eq. (\ref{AEq:UBEx:MinimumTau}) into Eqs. (\ref{Eq:SteSol:UBEx:WiTRelationFinal2}) with lower signs chosen  and (\ref{Eq:SteSol:UBEx:DeltaTRelation}), one obtains Eqs. (\ref{Eq:SteSol:UBEx:ExtensionalViscosityMinimumPosition}) and (\ref{Eq:SteSol:UBEx:ExtensionalViscosityMinimumValue}), respectively.
\subsection{
\label{SApp:Viscosities:PlaEx}
Monotonicity of the planar extensional viscosities}
Eliminating $\mathrm{Wi}$ from Eq. (\ref{Eq:SteSol:PlaEx:DeltaTRelation}) using Eq. (\ref{Eq:SteSol:PlaEx:WiTRelationFinal}), differentiating the result with respect to $\mathbb{T}$, and rearranging, one obtains
\begin{equation}
\label{AEq:PlaEx:DDeltaOneOverDTau}
\dfrac{\mathrm{d}\mathbb{N}_1}{\mathrm{d}\mathbb{T}}=
\dfrac{(1-2\epsilon)\mathbb{T}}{\sqrt{4\epsilon^2+(1-2\epsilon)\mathbb{T}^2}},
\end{equation}
which is clearly non-negative. Having applied the chain rule to $\mathbb{N}_1$, as in Appendix \ref{SApp:Viscosities:UBEx}, one demonstrates the non-negativity of $\mathrm{d} \mathbb{N}_1 / \mathrm{d}\mathrm{Wi}$. Then, one shows that $\mathrm{d} \mathbb{N}_2 / \mathrm{d}\mathrm{Wi}\geq 0$ by differentiating Eq. (\ref{Eq:SteSol:PlaEx:Delta2TDelta1Relation}) with respect to $\mathrm{Wi}$. Therefore, both planar extensional viscosities are increasing monotonically with $\mathrm{Wi}$.
\section{
\label{App:MainResult}
Derivation of the main results}
\subsection{
\label{SApp:MainResult:Shear}
Start-up of steady shear flow (trigonometric form $\mathfrak{T}$)}
To solve Eqs. (\ref{Eq:DFormulation:SteSh:1N1Evolution}) and (\ref{Eq:DFormulation:SteSh:2SEvolution}), one introduces the new variables, $\tilde{\mathbb{N}}_1$ and $\tilde{\mathbb{S}}$, which are the deviations of $\mathbb{N}_1^+$ and $\mathbb{S}^+$, respectively, from their steady-flow values,
\begin{align}
\tilde{\mathbb{N}}_1 &= \mathbb{N}_1-\mathbb{N}_1^+, \\
\tilde{\mathbb{S}}     &= \mathbb{S} - \mathbb{S}^+.
\end{align}
After this substutution, Eqs. (\ref{Eq:DFormulation:SteSh:1N1Evolution}) and (\ref{Eq:DFormulation:SteSh:2SEvolution}) become, respectively, 
\begin{align}
\tilde{\mathbb{N}}'_1 & = -(1 + 2\mathrm{Wi} \mathbb{N}_1)\tilde{\mathbb{N}}_1 + 2 \mathrm{Wi} \tilde{\mathbb{S}} + \mathrm{Wi} \tilde{\mathbb{N}}_1^2, \label{AEq:Solution:1N1Evolution}\\
\tilde{\mathbb{S}}' &= - (1 + \mathrm{Wi} \mathbb{N}_1)\tilde{\mathbb{S}} - \mathrm{Wi} \mathbb{S} \tilde{\mathbb{N}}_1 + \mathrm{Wi} \tilde{\mathbb{N}}_1 \tilde{\mathbb{S}}, \label{AEq:Solution:2SEvolution}
\end{align}
with the initial conditions $\tilde{\mathbb{N}}_1(0)=\mathbb{N}_1$ and $\tilde{\mathbb{S}}(0)=\mathbb{S}$. Subtracting Eq. (\ref{AEq:Solution:2SEvolution}) multiplied by $\tilde{\mathbb{N}}_1$ from Eq. (\ref{AEq:Solution:1N1Evolution}) multiplied by $\tilde{\mathbb{S}}$,  dividing the result by $\tilde{\mathbb{S}}^2$, and introducing $V(\bar{t})=\tilde{\mathbb{N}}_1/\tilde{\mathbb{S}}$ yield
\begin{equation}
V' = 2\mathrm{Wi} -\mathrm{Wi} \mathbb{N}_1 V + \mathrm{Wi} \mathbb{S} V^2, \label{AEq:Solution:Riccati}
\end{equation}
with $V(0)=\mathbb{N}_1/\mathbb{S}$. This is a Riccati differential equation, which can be solved by standard analytical methods.\cite{Ince2006} The solution of Eq. (\ref{AEq:Solution:Riccati}) can be written as
\begin{equation}
V(\bar{t}) =\dfrac{1}{2\mathbb{S}}\left(\mathbb{N}_1+\sqrt{\Delta}\dfrac{\sin \omega \bar{t}+\dfrac{\mathbb{N}_1}{\sqrt{\Delta}} \cos \omega \bar{t}  }{\cos \omega \bar{t} - \dfrac{\mathbb{N}_1}{\sqrt{\Delta}} \sin \omega \bar{t} }\right), \label{AEq:Solution:V}
\end{equation}
where $\Delta=-\mathbb{N}_1^2+8\mathbb{S}$ (see column I of Table \ref{Tab:FormFunctions}) and $\omega = \mathrm{Wi} \sqrt{\Delta}/2$. It should be noted that $\Delta>0$, meaning that all functions appearing in Eq. (\ref{AEq:Solution:V}) are real-valued. The positivity of $\Delta$ can be shown by the following transformations:
\begin{equation}
\Delta=-\mathbb{N}_1^2+8\mathbb{S} = -\dfrac{4}{\epsilon^2}\mathrm{Wi}^2\mathbb{S}^4+8\mathbb{S} = -\dfrac{2}{\epsilon}(\epsilon-\mathbb{S})\mathbb{S}+8\mathbb{S} = 2\mathbb{S} \left(3+\dfrac{\mathbb{S}}{\epsilon} \right)>0,
\end{equation}
where Eqs. (\ref{Eq:SteSol:Sh:Delta1SigmaRelation}) and (\ref{Eq:SteSol:Sh:WiSigmaRelationFinal}) were used subsequently.
\par 
Then, the relation between $\tilde{\mathbb{N}}_1$ and $\tilde{\mathbb{S}}$,
\begin{equation}
\tilde{\mathbb{N}}_1=V(\bar{t})\tilde{\mathbb{S}}, \label{AEq:Solution:NSRelation}
\end{equation}
is used to eliminate $\tilde{\mathbb{N}}_1$ from Eq. (\ref{AEq:Solution:2SEvolution}); this leads to
\begin{equation}
\tilde{\mathbb{S}}' = - \left[1 + \mathrm{Wi} \mathbb{N}_1+\mathrm{Wi}\mathbb{S} V(\bar{t})\right]\tilde{\mathbb{S}} + \mathrm{Wi} V(\bar{t}) \tilde{\mathbb{S}}^2, \label{AEq:Solution:Bernoulli}
\end{equation}
with $\tilde{\mathbb{S}}(0)=\mathbb{S}$ and $V(\bar{t})$ given by Eq. (\ref{AEq:Solution:V}). Equation (\ref{AEq:Solution:Bernoulli}) is a Bernoulli differential equation; having solved it using the standard analytical  methods,\cite{Ince2006} one obtains
\begin{equation}
\tilde{\mathbb{S}} = \dfrac{\mathbb{S} K \left(\cos \omega \bar{t} -\dfrac{\mathbb{N}_1}{\sqrt{\Delta}}\sin \omega \bar{t} \right)}{C \exp \Omega \bar{t}+\mathrm{Wi} (A \cos \omega \bar{t} + B \sin \omega \bar{t} )}, \label{AEq:Solution:SHat}
\end{equation}
where $\Omega$, $K$, $C$, $A$, and $B$ are given in column I of Table \ref{Tab:FormFunctions}.
Finally, substituting Eqs. (\ref{AEq:Solution:V}) and (\ref{AEq:Solution:SHat}) into Eq. (\ref{AEq:Solution:NSRelation}) and performing the multiplication, one arrives at
\begin{equation}
\tilde{\mathbb{N}}_1 = \dfrac{\mathbb{N}_1 K \left(\cos \omega \bar{t} + \dfrac{\mathbb{N}_1 + 2 \mathrm{Wi} \mathbb{S}}{\mathrm{Wi} \mathbb{N}_1 \sqrt{\Delta}}\sin \omega \bar{t}\right)}{C \exp \Omega \bar{t}+\mathrm{Wi} (A \cos \omega \bar{t} + B \sin \omega \bar{t} )}. \label{AEq:Solution:N1Hat}
\end{equation}
It is seen that Eqs. (\ref{AEq:Solution:SHat}) and (\ref{AEq:Solution:N1Hat}) are of very similar form, the main difference being the coefficient in front of $\sin \omega \bar{t}$ in the numerator. This coefficient is denoted $a$; the expressions specifying $a$ corresponding to different material functions are found in Table \ref{Tab:a}.
\par 
Reverting to the original variables, $(\mathbb{S}^+, \mathbb{N}_1^+)$, and then expressing the normalized material functions as described in Table \ref{Tab:MatFunctionsVsDimlessVars}, one obtains the main result for start-up of steady shear flow: the trigonometric form ($\mathfrak{T}$) presented in Table \ref{Tab:Forms}.
\subsection{
\label{SApp:MainResult:Extensional}
Start-up of steady extensional flows (hyperbolic form $\mathfrak{H}$)}
Derivation of the main result for start-up of steady uniaxial, biaxial, and planar extensional flows is similar to that for start-up of steady shear flows and goes through the same steps; therefore, only the key points of the derivation shall be given here.
\par 
The new variables, $(\tilde{\mathbb{T}},\tilde{\mathbb{N}}_1)$, are introduced as the deviations of the old variables, $(\mathbb{T}^+,\mathbb{N}_1^+)$, from their steady-flow values,
\begin{align}
\tilde{\mathbb{T}} &= \mathbb{T}-\mathbb{T}^+, \\
\tilde{\mathbb{N}}_1     &= \mathbb{N}_1 - \mathbb{N}_1^+.
\end{align}
With this, Eqs. (\ref{Eq:DFormulation:UBEx:1TEvolution}) and (\ref{Eq:DFormulation:UBEx:2N1Evolution}) become
\begin{align}
\tilde{\mathbb{T}}' &= - (1+2\mathrm{Wi} \mathbb{T})\tilde{\mathbb{T}} + 2 \mathrm{Wi} \tilde{\mathbb{N}}_1 + \mathrm{Wi} \tilde{\mathbb{T}}^2, \\
\tilde{\mathbb{N}}'_1 &= - (1+\mathrm{Wi} \mathbb{T}) \tilde{\mathbb{N}}_1 - \mathrm{Wi} \mathbb{N}_1 \tilde{\mathbb{T}} + \mathrm{Wi} (\tilde{\mathbb{T}} \pm \tilde{\mathbb{N}}_1) + \mathrm{Wi} \tilde{\mathbb{T}}\tilde{\mathbb{N}}_1, \label{AEq:SolutionB:2N1Evolution-UBEx}
\end{align}
respectively, while Eqs. (\ref{Eq:DFormulation:PlaEx:1TEvolution}) and (\ref{Eq:DFormulation:PlaEx:2N1Evolution}) become
\begin{align}
\tilde{\mathbb{T}}' &= - (1+2\mathrm{Wi} \mathbb{T})\tilde{\mathbb{T}} + 2 \mathrm{Wi} \tilde{\mathbb{N}}_1 + \mathrm{Wi} \tilde{\mathbb{T}}^2, \\
\tilde{\mathbb{N}}'_1 &= - (1+\mathrm{Wi} \mathbb{T}) \tilde{\mathbb{N}}_1 - \mathrm{Wi} \mathbb{N}_1 \tilde{\mathbb{T}} + 2 \mathrm{Wi} \tilde{\mathbb{T}} + \mathrm{Wi} \tilde{\mathbb{T}} \tilde{\mathbb{N}}_1, \label{AEq:SolutionB:2N1Evolution-PlaEx}
\end{align}
respectively, the initial conditions being $\tilde{\mathbb{T}}(0)=\mathbb{T}$, $\tilde{\mathbb{N}}_1(0)=\mathbb{N}_1$ in both cases. Then, a new variable, $V(\bar{t})$, is defined by
\begin{equation}
\label{AEq:SolutionB:TNRelation}
\tilde{\mathbb{T}} = V(\bar{t}) \tilde{\mathbb{N}}_1,
\end{equation}
and a Riccati equation for $V$ is constructed, as shown in Appendix \ref{SApp:MainResult:Shear}. For start-up of uniaxial and biaxial extensional flows, this equation is
\begin{equation}
\label{AEq:SolutionB:Riccati-UBEx}
V' =2\mathrm{Wi} -\mathrm{Wi} (\mathbb{T} \pm 1)V + \mathrm{Wi} (\mathbb{N}_1-1)V^2,
\end{equation}
while for start-up of planar extensional flows,
\begin{equation}
\label{AEq:SolutionB:Riccati-PlaEx}
V' = 2\mathrm{Wi} -\mathrm{Wi} \mathbb{T} V + \mathrm{Wi} (\mathbb{N}_1-2)V^2,
\end{equation}
with $V(0)=\mathbb{T}/\mathbb{N}_1$ in both cases.
\par 
The solution of Eq. (\ref{AEq:SolutionB:Riccati-UBEx}) is
\begin{equation}
\label{AEq:SolutionB:V-UBEx}
V = \dfrac{1}{2(\mathbb{N}_1-1)}\left(\mathbb{T} \pm 1 - \sqrt{\Delta} \dfrac{\sinh \omega \bar{t} + \dfrac{ \pm \mathbb{N}_1 +2\mathbb{T} - \mathbb{N}_1 \mathbb{T}}{\mathbb{N}_1 \sqrt{\Delta}} \cosh \omega \bar{t}}{\cosh \omega \bar{t} + \dfrac{ \pm \mathbb{N}_1 +2\mathbb{T} - \mathbb{N}_1 \mathbb{T}}{\mathbb{N}_1 \sqrt{\Delta}} \sinh \omega \bar{t}} \right),
\end{equation}
with $\Delta=9-8\mathbb{N}_1\pm 2 \mathbb{T}+\mathbb{T}^2>0$ (see column II of Table \ref{Tab:FormFunctions}), while the solution of Eq. (\ref{AEq:SolutionB:Riccati-PlaEx}) is
\begin{equation}
\label{AEq:SolutionB:V-PlaEx}
V  = \dfrac{1}{2(\mathbb{N}_1-2)} \left(\mathbb{T} - \sqrt{\Delta} \dfrac{\sinh \omega \bar{t} + \dfrac{(4-\mathbb{N}_1)\mathbb{T}}{\mathbb{N}_1 \sqrt{\Delta}} \cosh \omega \bar{t}}{\cosh \omega \bar{t} + \dfrac{(4-\mathbb{N}_1)\mathbb{T}}{\mathbb{N}_1 \sqrt{\Delta}} \sinh \omega \bar{t}} \right),
\end{equation}
with $\Delta=16-8\mathbb{N}_1+\mathbb{T}^2>0$ (see column III of Table \ref{Tab:FormFunctions}). In both cases, $\omega = \mathrm{Wi} \sqrt{\Delta}/2$.
\par 
These solutions, together with Eq. (\ref{AEq:SolutionB:TNRelation}), are used to eliminate $\tilde{\mathbb{T}}$ from Eqs. (\ref{AEq:SolutionB:2N1Evolution-UBEx}) and (\ref{AEq:SolutionB:2N1Evolution-PlaEx}), respectively. The resulting Bernoulli equations are solved, yielding $\tilde{\mathbb{N}}_1$. Then, $\tilde{\mathbb{T}}$ is found from Eq. (\ref{AEq:SolutionB:TNRelation}). Finally, for the planar extension case, $\mathbb{N}_2^+$ is obtained using Eq. (\ref{Eq:DFormulation:PlaEx:3N2Evolution}) after reverting to the original variables, ($\mathbb{T}^+,\mathbb{N}_1^+$). Calculating the normalized material functions according to Table \ref{Tab:MatFunctionsVsDimlessVars} leads to the hyperbolic form $\mathfrak{H}$ presented in Table \ref{Tab:Forms}.
\subsection{
\label{SApp:MainResult:Cessation}
Cessation of steady shear and extensional flows (exponential form $\mathfrak{E}$)}
The derivation of the exact analytical solutions for the cessation case is trivial compared to the start-up case. Separating the variables in Eq. (\ref{Eq:DFormulation:Cessation:GeneralEquation}) leads to 
\begin{equation}
\dfrac{\mathrm{d}(1+\mathrm{Wi} \mathbb{X}^-) } {1+\mathrm{Wi} \mathbb{X}^-}-\dfrac{\mathrm{d}\mathbb{X}^-}{\mathbb{X}^-}=\mathrm{d}\bar{t}.
\end{equation}
Integrating this with the initial condition $\mathbb{X}^-(0)=\mathbb{X}$ yields
\begin{equation}
\mathbb{X}^- = \dfrac{\mathbb{X}}{(1+\mathrm{Wi} \mathbb{X}) \exp \bar{t}-\mathrm{Wi} \mathbb{X}}.
\end{equation}
Combining this with Eqs. (\ref{Eq:DFormulation:Cessation:MatFunShear}) and (\ref{Eq:DFormulation:Cessation:MatFunExtensional}), one obtains the exponential form $\mathfrak{E}$ presented in Table \ref{Tab:Forms}.
\end{appendix}

\section*{Data availability}
Data sharing is not applicable to this article as no new data were created or analyzed in this study.

\bibliography{SLPTT}

\end{document}